\documentclass[draftcls, onecolumn]{IEEEtran}
\usepackage{amsmath,amssymb}
\usepackage{graphicx}
\usepackage{mathrsfs}
\usepackage{color}
\usepackage{tikz}
\usepackage{booktabs}
\usepackage{enumerate}
\usepackage{psfrag}	
\usepackage{array}
\usepackage{multirow,hhline}
\usepackage{exscale}
\usepackage{colortbl}
\usepackage{epsfig,subfigure}
\usepackage{cite}
\usepackage{amsthm}
\usepackage[pdftex,bookmarks=true]{hyperref}
\usepackage[]{algorithm2e,algorithmicx}

\usepackage[outline]{contour}
\contourlength{1.2pt}
\usepackage{tikz}
\usepackage{pgfplots}
\usepackage{pgfplotstable}
\usepackage{rotate}
\usepgfplotslibrary{units}
\usetikzlibrary{%
patterns,%
calc,%
fit,%
arrows,%
plotmarks,%
shadows,%
chains,%
shapes%
}
\usepackage[outline]{contour}
\contourlength{1.2pt}
\tikzset{>=latex'} 
\tikzstyle{every picture}+=[remember picture] 
\definecolor{rubgray}{cmyk}{0.03,0.03,0.03,0.1}

\newtheorem{theorem}{Theorem}
\newtheorem{lemma}{Lemma}
\newtheorem{remark}{Remark}

\newtheorem{definition}{Definition}

\newcommand{\mat}[1]{\ensuremath{\boldsymbol{#1}}}
\renewcommand{\vec}[1]{\ensuremath{\boldsymbol{#1}}}
\newcommand{\y}[0]{\ensuremath{\boldsymbol{y}}}
\newcommand{\p}[0]{\ensuremath{\boldsymbol{p}}}
\newcommand{\z}[0]{\ensuremath{\boldsymbol{z}}}
\renewcommand{\H}[0]{\ensuremath{\boldsymbol{H}}}
\newcommand{\x}[0]{\ensuremath{\boldsymbol{x}}}
\newcommand{\X}[0]{\ensuremath{\boldsymbol{X}}}

\renewcommand{\c}[0]{\ensuremath{\boldsymbol{c}}}
\newcommand{\e}[0]{\ensuremath{\boldsymbol{e}}}
\newcommand{\U}[0]{\ensuremath{\boldsymbol{U}}}
\newcommand{\V}[0]{\ensuremath{\boldsymbol{V}}}
\newcommand{\I}[0]{\ensuremath{\boldsymbol{I}}}
\newcommand{\D}[0]{\ensuremath{\boldsymbol{D}}}
\renewcommand{\u}[0]{\ensuremath{\boldsymbol{u}}}
\renewcommand{\v}[0]{\ensuremath{\boldsymbol{v}}}

\renewcommand{\d}[0]{\ensuremath{\boldsymbol{d}}}

\newcommand{\W}[0]{\ensuremath{\boldsymbol{W}}}
\newcommand{\ind}[0]{\ensuremath{\mathsf{I}}}

\DeclareMathOperator{\tr}{trace}

\definecolor{Ogreen}{rgb}{.2,.4,.2}

\graphicspath{{./fig/}}

\begin{document}

\IEEEoverridecommandlockouts
\title{Cyclic Communication and the Inseparability of MIMO Multi-way Relay Channels}
\author{
\IEEEauthorblockN{Anas~Chaaban and Aydin Sezgin}\\
\IEEEauthorblockA{}
\thanks{%
A. Chaaban is with the Division of Computer, Electrical, and Mathematical Sciences and Engineering, King Abdullah University of Science and Technology, Thuwal, Saudi Arabia. Email: anas.chaaban@kaust.edu.sa.

A. Sezgin is with the Institute of Digital Communication Systems, Ruhr-Universit\"at Bochum (RUB), Germany. Email: aydin.sezgin@rub.de.

Part of this work has been presented in the European Wireless conference 2014 \cite{ChaabanOchsSezgin_EW}.
}
}

\maketitle

\begin{abstract}
The $K$-user MIMO multi-way relay channel (Y-channel) consisting of $K$ users with $M$ antennas each and a common relay node with $N$ antennas is studied in this paper. Each user wants to exchange messages with all the other users via the relay. A transmission strategy is proposed for this channel. The proposed strategy is based on two steps: channel diagonalization and cyclic communication. The channel diagonalization is applied by using zero-forcing beam-forming. After channel diagonalization, the channel is decomposed into parallel sub-channels. Cyclic communication is then applied, where signal-space alignment for network-coding is used over each sub-channel. The proposed strategy achieves the optimal DoF region of the channel if $N\leq M$. To prove this, a new degrees-of-freedom outer bound is derived. As a by-product, we conclude that the MIMO Y-channel is not separable, i.e., independent coding on separate sub-channels is not enough, and one has to code jointly over several sub-channels.
\end{abstract}


\section{Introduction}
Experts have predicted that the number of devices with communication capability will rise to 50 billions by 2020 \cite{Evans_IoT}. The resulting web of devices connected by the Internet-of-Things (IoT) and Machine-to-Machine (M2M) communications for instance will lead to more sophisticated network topologies. Communication over such networks is in general multi-way, where communicating pairs of nodes exchange information in both directions such as in the two-way channel \cite{Shannon_TWC,Han}.

Beside multi-way communication, a key aspect of future networks is relaying which can play a key role in improving transmission rates. In multi-way networks in particular, the potential of multi-way relaying can be of great importance \cite{ChaabanSezginFnT}. This is especially true in scenarios where physical-layer network coding can be applied, which can significantly boost the performance of a network \cite{WilsonNarayananPfisterSprintson, AvestimehrSezginTse}.

For the aforementioned reasons, the multi-way relay channel (MWRC) which combines both aspects (multi-way and relaying) is an integral part of future networks. The MWRC consists of multiple users that want to exchange information via a common relay node. In its simplest form with two users, we get the so called two-way relay channel TWRC. The TWRC is a fundamental scenario that has been introduced in \cite{RankovWittneben}, and studied thoroughly recently in \cite{KimDevroyeMitranTarokh, GunduzTuncelNayak, OechteringBjelakovicSchnurrBoche, NamChungLee_IT, AlsharoaGhazzaiAlouini, ShaqfehZafarAlnuweiriAlouini}. Several transmission strategies for the TWRC including compress-forward and lattice coding have been examined lately, leading to the capacity of the TWRC within a constant gap \cite{AvestimehrSezginTse, NamChungLee_IT}.

Although the TWRC has become well-understood recently, the MWRC has not reached a similar status yet, although several researches have focused on this network recently. For instance, \cite{NgoLarsson, SezginAvestimehrKhajehnejadHassibi, SezginBocheAvestimehr} study the multi-pair TWRC, \cite{MokhtarMohassebNafieElGamal, OngKellettJohnson_IT, GunduzYenerGoldsmithPoor_IT} study the multi-cast MWRC, \cite{MatthiesenZapponeJorswieck} studies the MWRC with cyclic message exchange, and \cite{LeeLimChun, ChenweiWang, ChaabanSezginAvestimehr_YC_SC} study the MWRC with multiple uni-cast message exchange. In this paper, we focus on the latter variant of the MWRC, i.e., the MWRC with multiple uni-cast message exchange, also known as the Y-channel \cite{LeeLim}.

In the $K$-user Y-channel, several users want to exchange information in all directions via the relay. In particular, user $i\in\{1,\cdots,K\}$ wants to communicate with user $j\in\{1,\cdots,K\}\setminus\{i\}$. The extension of the TWRC to the Y-channel is not straightforward, and many challenges have to be tackled when making this step. One of the challenges is in deriving capacity upper bounds. While the capacity of the TWRC can be approximated with a high-precision using the cut-set bounds \cite{CoverThomas}, the capacity of the $K$-user Y-channel requires new bounds. Such bounds have been derived in \cite{ChaabanSezginAvestimehr_YC_SC, ChenweiWang}. Another challenge is in finding the best communication strategy. The $K$-user Y-channel requires, in addition to bi-directional communication strategies used in the TWRC, more involved strategies such as cyclic communication \cite{ChaabanSezginISIT, ChaabanSezgin_ISIT12_Y} and detour schemes \cite{ZewailMohassebNafieElGamal}. 

The Single-Input Single-Output (SISO) $K$-user Y-channel has been studied in \cite{ChaabanSezginAvestimehr_YC_SC}. Here, we focus on the Multiple-Input Multiple-Output (MIMO) case. The MIMO Y-channel has been initially introduced in \cite{LeeLim}, where the strategy of signal-space alignment for network-coding was used. In their paper, Lee {\it et al.} characterized the optimal sum degrees-of-freedom (DoF) of the 3-user MIMO Y-channel under some conditions on the ratio of the number of antennas at the users and the relay. However, a complete sum-DoF characterization of the general 3-user MIMO Y-channel was not available until~\cite{ChaabanOchsSezgin} where a novel upper bound and a general transmission strategy were developed, thus settling this problem. The MIMO Y-channel with more than 3 users has also been studied in \cite{TianYenerMIMOMW,LeeLeeLee}. In \cite{TianYenerMIMOMW}, Tian and Yener have studied the multi-cluster MIMO Y-channel and characterized the sum-DoF of the channel under some conditions on the number of antennas, while in \cite{LeeLeeLee}, Lee {\it et al.} proposed a transmission strategy for the $K$-user MIMO Y-channel and derived its achievable DoF. Despite the intensive work on the Y-channel, many questions remain open. For instance, the sum-DoF of the general $K$-user MIMO Y-channel remains open to date. A recent development on this front has been achieved recently, when Wang has characterized the sum-DoF of the 4-user MIMO Y-channel in \cite{ChenweiWang}. Another question is on the DoF region of the MIMO Y-channel which is still unknown. Recently, the DoF region of the 3-user and 4-user cases was studied in \cite{ZewailNafieMohassebGamal}.

The importance of the DoF region is that it reflects the trade-off between the achievable different DoF of different users, contrary to the sum-DoF which does not. This trade-off is essential in cases where the DoF demand by different users is not the same, such as in a network with prioritized users. In such cases, it is interesting to know what is the maximum DoF that can be achieved by some users under some constraints on the DoF of other users. This question can be answered by the DoF region. Also, by obtaining the DoF region, the sum-DoF is obtained as a by-product.

In this paper, we focus on the DoF {\it region} of the $K$-user MIMO Y-channel. We develop a communication strategy for the $K$-user MIMO Y-channel with $M$ antennas at the users, and $N\leq M$ antennas at the relay. This case models a situation where it is easier to mount antennas at the users than at the relay node, such as when the relay is a satellite node. Our proposed strategy revolves around two ideas: (i) channel diagonalization and (ii) cyclic communication using physical-layer network-coding. Channel diagonalization is performed by zero-forcing beam-forming~\cite{Jindal} using the Moore-Penrose pseudo-inverse. After channel diagonalization, the MIMO Y-channel is decomposed into a set of parallel SISO Y-channels (sub-channels). Then, cyclic communication is performed over these sub-channels. A cyclic communication strategy ensures information exchange over a set of users in a cyclic manner, such as exchanging a signal from user $1$ to $2$, $2$ to $3$, and $3$ to $1$ thus constituting the cycle $1\to 2\to 3\to 1$. In cyclic communication, the users send a set of symbols to the relay, which decodes functions of these symbols \cite{NazerGastpar} and forwards these functions to the users. These functions have to be designed appropriately, so that each user can extract his desired symbol from these functions after reception. Note that the $K$-user Y-channel has cycles of length 2 ($1\to 2\to 1$ e.g.) up to length $K$ ($1\to2\to\cdots\to K\to1$ e.g.). We call the transmission strategy corresponding to an $\ell$-cycle (cycle of length $\ell$) an $\ell$-cyclic strategy. The efficiency of the proposed $\ell$-cyclic strategy is $(\ell+1)/\ell$ symbol/sub-channel (or DoF/dimension).

Note that after channel diagonalization, the channel has similarities to the linear-deterministic 3-user SISO Y-channel studied in \cite{ChaabanSezgin_YC_Reg} which is a set of parallel binary Y-channels, some of which are not fully-connected. The difference is that the parallel SISO Y-channels obtained after diagonalization of the MIMO Y-channel are complex-valued. Furthermore, the work in \cite{ChaabanSezgin_YC_Reg} considers only the 3-user case, and the extension to the $K$-user case is not considered. Thus, the main difference between this work and the one in \cite{ChaabanSezgin_YC_Reg} is that here we:
\begin{enumerate}
\item extend the scheme to the complex-valued channel with $K\geq3$ user,
\item provide a graphical illustration of the problem in the form of a message flow graph, 
\item show that with $K$ users, cyclic communication over cycles of various lengths has to be considered, and
\item propose an optimal resource allocation strategy which distributes the streams to be communicated over the available sub-channels, and uses the optimal strategies over these sub-channels. 
\end{enumerate}

The question that arises at this point is: Is it optimal to treat each sub-channel of the MIMO Y-channel separately \cite{CadambeJafar_Inseperability}? Or is it better to encode jointly over sub-channels? To answer this question, one has to optimize the transmission strategy, and observe if the optimized solution requires joint encoding over spatial-dimensions. With this goal in mind, we propose a resource allocation that allocates sub-channels to cyclic strategies based on their efficiencies. The proposed resource allocation is proved to be optimal by deriving a DoF region outer bound using a genie-aided approach. Similar to \cite{ChaabanOchsSezgin}, the derived genie-aided bound converts the Y-channel into a MIMO point-to-point channel whose DoF is known \cite{Telatar}. As a result, a DoF region characterization for the $K$-user MIMO Y-channel with $N\leq M$ is obtained. This provides {\it the first DoF region characterization for the $K$-user MIMO Y-channel.}

With the optimal strategy at hand, we can go back to the channel separability question. We observe that the DoF-region-optimal strategy for the MIMO Y-channel treats the parallel sub-channels jointly, where encoding over spatial dimensions is necessary. We conclude that the MIMO Y-channel is not separable. However, from sum-DoF point-of-view (instead of DoF-region), separate encoding over each sub-channel is optimal. Another interesting observation is that the optimal strategy is in fact a combination of different cyclic strategies with different efficiencies. In other words, it is not enough to rely on the cyclic strategy with highest efficiency, i.e., the 2-cyclic strategy.

In the next section, we formally define the $K$-user MIMO Y-channel. We introduce the main result of the paper, which is a DoF region characterization of the case $N\leq M$ in Section \ref{SecMainResult}. Next, we introduce our communication strategy by using a toy-example consisting of a 3-user Y-channel in Section \ref{Sec:3UserYChannel}. The communication strategy for the $K$-user case is described in detail in Section \ref{Sec:Achievability}. Comments on the regime where $N>M$ and on the inseparability of the Y-channel are given in Sections \ref{Sec:NgM} and \ref{Sec:SubChannelSep}, respectively. Finally, we conclude the paper with a discussion in Section \ref{Sec:Conclusion}.

\begin{figure*}[t]
\centering
\subfigure[Uplink.]{
{\includegraphics[width=.42\textwidth]{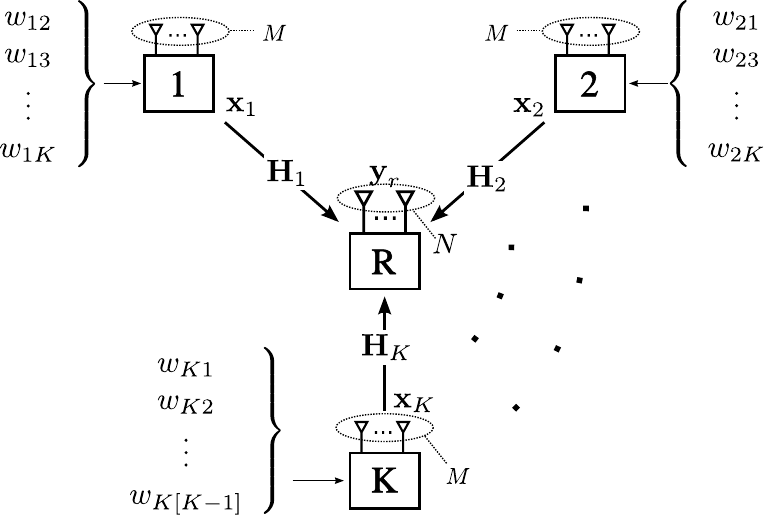}}
\label{Fig:ModelU}
}
\hspace{1cm}
\subfigure[Downlink.]{
{\includegraphics[width=.42\textwidth]{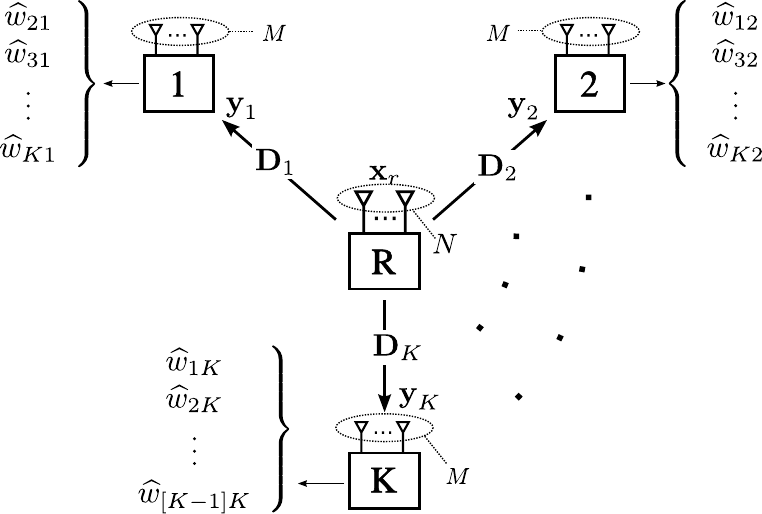}}
\label{Fig:ModelD}
}
\caption{The $K$-user MIMO Y-channel in the uplink and downlink. Each user $i\in\{1,\cdots,K\}$ sends $K-1$ messages $m_{ij}$, $j\in\{1,\cdots,K\}\setminus\{i\}$ where $m_{ij}$ is intended to user $j$. Consequently, each user decodes $K-1$ messages.}
\label{Fig:Ychannel}
\end{figure*}

\section{Notation and System Model}
\label{Sec:Model}
\subsection{Notation}
The following notation is used throughout the paper. We use bold-face lower-case ($\x$) and upper-case ($\X$) letters to denote vectors and matrices, respectively, and we use normal fonts ($x$) and calligraphic fonts ($\mathcal{X}$) to denote scalars and sets, respectively.  We denote the $N\times N$ identity matrix and the $q\times 1$ zero vector by $\I_N$ and $\vec{0}_q$, respectively. We say that $\x\sim\mathcal{CN}(\vec{m},\mat{Q})$ when $\x$ is a complex Gaussian random vector with mean $\vec{m}$ and covariance matrix $\mat{Q}$. We use $\X^{H}$ and $\X^{-1}$ to denote the Hermitian transpose and the inverse of a matrix $\X$, respectively. We also use $\x^{\tau}$ to denote the length-$\tau$ sequence $(\x(1),\cdots,\x(\tau))$. A sequence $\x^\tau$ is i.i.d. if its components are independent and identically distributed. The function $\ind_\mathcal{A}(x)$ is an indicator function which returns 1 if $x\in\mathcal{A}$ and 0 otherwise, and $\bar{\ind}_\mathcal{A}(x)$ is the inverse indicator function.

\subsection{System Model}
The $K$-user MIMO Y-channel consists of $K$ users which want to establish full message-exchange via a relay as shown in Figures \ref{Fig:ModelU} and \ref{Fig:ModelD}. All nodes are assumed to be full-duplex 
with power $\rho$.\footnote{Note that an equal power constraint $\rho$ can be assumed without loss of generality, since different powers can be incorporated into the channel.} The relay has $N$ antennas, and the users are assumed to be identical in terms of the number of antennas, with $M$ antennas at each user. User $i\in\mathcal{K}=\{1,\cdots,K\}$ has a message $w_{ij}$ to be sent to user $j$ for all $j\in\mathcal{K}\setminus\{i\}$. The message $w_{ij}$ is a realization of a random variable $W_{ij}$ uniformly distributed over the set $\mathcal{W}_{ij}=\{1,\cdots,2^{\tau R_{ij}(\rho)}\}$ where $R_{ij}(\rho)>0$ denotes the rate of the message, and $\tau$ denotes the number of transmissions (channel uses).

At time instant $t\in\{1,\cdots,\tau\}$, user $i$ sends $\x_{i}(t)\in\mathbb{C}^M$ which is a codeword symbol constructed from the messages $w_{ij}$, $j\neq i$, and from $\y_i^{t-1}$, the received signals of user $i$ up to time instant $t-1$. This transmit signal has to satisfy the power constraint, i.e., 
\begin{align}
\tr(\mathbb{E}[\x_i\x_i^{H}])&\leq \rho.
\end{align}
The received signal at the relay is given by (cf. Figure \ref{Fig:ModelU})
\begin{align}
\y_{r}(t)=\sum_{i=1}^K\H_i\x_i(t)+\z_{r}(t),
\end{align}
which is an $N\times1$ vector, where the noise $\z_{r}(t)\sim\mathcal{CN}(\vec{0}_N,\vec{I}_N)$ is i.i.d. over time. Here $\H_i$ is the $N\times M$ complex channel matrix from user $i$ to the relay, which is assumed to be constant throughout the $\tau$ channel uses, and has rank $\min\{M,N\}$. The relay transmit signal at time $t$ is denoted $\x_{r}(t)\in\mathbb{C}^N$, it satisfies
\begin{align}
\tr(\mathbb{E}[\x_r\x_r^{H}])&\leq \rho,
\end{align}
and it is constructed from $\y_r^{t-1}$, the received signal at the relay up to time instant $t-1$. The received signal at user $i$ is given by (cf. Fig. \ref{Fig:ModelD})
\begin{align}
\label{ReceivedSignal}
\y_{i}(t)=\D_i\x_{r}(t)+\z_{i}(t),
\end{align}
which is an $M\times1$ vector, where the noise $\z_{i}(t)\sim\mathcal{CN}(\vec{0}_M,\vec{I}_M)$ is i.i.d. over time\footnote{The time index $t$ will be suppressed henceforth.}, and $\D_i$ is the $M\times N$ downlink constant complex channel matrix from the relay to user $i$, and has rank $\min\{M,N\}$. After $\tau$ channel uses, user $i$ has $\y_i^\tau$ from which it tries to decode $w_{ji}$, $j\neq i$, by using its messages $w_{ij}$ as side information. After decoding, it obtains $\hat{w}_{ji}$, $j\neq i$. An error occurs if $w_{ji}\neq \hat{w}_{ji}$ for some distinct $i,j\in\mathcal{K}$.

A rate $R_{ij}(\rho)$ is said to be achievable if there exist a strategy (encoding and decoding strategies) that provides an error probability $\text{Prob}[w_{ij}\neq \hat{w}_{ij}]$ that vanishes as $\tau\to\infty$. The DoF of the corresponding message $w_{ij}$ is defined as \cite{CadambeJafar_KUserIC}
\begin{align}
\label{DoFDef}
d_{ij}=\lim_{\substack{\rho\to\infty}}\frac{R_{ij}(\rho)}{\log(\rho)},
\end{align}
and is said to be achievable if the corresponding $R_{ij}(\rho)$ satisfying \eqref{DoFDef} is. Let us collect the DoF of all messages in a DoF tuple $\d\in\mathbb{R}^{K(K-1)}$ defined as
\begin{align}
\label{Defpfvecd}
\d=(d_{12},\cdots,d_{1K},d_{21},d_{23},\cdots,d_{2K},\cdots,d_{K1},\cdots,d_{K[K-1]}).
\end{align}
A DoF tuple $\d$ is said to be achievable if its components are simultaneously achievable. We define the DoF region of the $K$-user Y-channel $\mathcal{D}_K$ as the set of all achievable DoF tuples $\d$. We also define the sum-DoF $d_\Sigma$ of the channel as the maximum achievable total DoF given by $$d_\Sigma=\max_{\d\in\mathcal{D}_K}(\d\cdot\boldsymbol{1}),$$
where $\boldsymbol{1}$ is a $K(K-1)\times 1$ vector of all ones.

Having defined the $K$-user MIMO Y-channel, we are ready to present the main result of the paper given in the next section.

\section{Main result}
\label{SecMainResult}
The main result of the paper is a characterization of the DoF region of the $K$-user MIMO Y-channel with $N\leq M$ as given in the following theorem.
\begin{theorem}
\label{Thm:DoFKUsers}
The DoF region $\mathcal{D}_K$ of the $K$-user MIMO Y-channel with $N\leq M$ is given by the set of tuples $\d\in\mathbb{R}^{K(K-1)}$ satisfying 
\begin{align}
\label{Eq:DoFBoundThm}
\sum_{i=1}^{K-1}\sum_{j=i+1}^{K}d_{p_ip_j}\leq N, \quad \forall \mathbf{p}
\end{align}
where $\mathbf{p}$ is a permutation of $(1,\cdots,K)$ and $p_i$ is its $i$-th component.
\end{theorem}
To show that no DoF tuple outside $\mathcal{D}_K$ is achievable, we derive a DoF upper bound based on a genie-aided approach that transforms the MIMO Y-channel into an $N\times (K-1)M$ MIMO point-to-point channel \cite{Telatar}. This upper bound leads to a DoF region outer bound that coincides with \eqref{Eq:DoFBoundThm} which proves the converse of Theorem \ref{Thm:DoFKUsers}. Details are given in Appendix \ref{Sec:Converse}. 

The achievability of this theorem is the main focus of the rest of the paper. The achievability is proved using three steps:
\begin{enumerate}
\item First, we use zero-forcing pre-coding (beam-forming) in the uplink, and zero-forcing post-coding in the downlink to diagonalize the channel, thus transforming it into a set of parallel SISO $K$-user Y-channels (sub-channels). This is described in Section \ref{Sec:ChDiag}.
\item Second, we perform physical-layer network coding over these sub-channels using different transmission strategies for different modes of message exchange. This step is explained in Section \ref{Sec:TranStra}.
\item Third, we solve a resource allocation problem that distributes the available sub-channels optimally among these strategies in Section \ref{Sec:ResAlloc}.
\end{enumerate}
By using this approach, we are able to show that any DoF tuple $\d$ in the DoF region $\mathcal{D}_K$ is achievable. 

It turns out that the optimal strategy for this channel requires encoding jointly over multiple sub-channels, and that it is not sufficient to encode over each sub-channel separately. Thus, as a by-product of this result, we conclude that the MIMO Y-channel with $N\leq M$ is inseparable \cite{CadambeJafar_Inseperability}. This aspect is elaborated in Section \ref{Sec:SubChannelSep} after proving the achievability of $\mathcal{D}_K$. But before we proceed, let us consider a toy-example with 3 users to illustrate the transmission strategies which achieve this outer bound.

\section{The 3-user Y-channel}
\label{Sec:3UserYChannel}
Here, we provide an informal preview of the achievability proof of Theorem \ref{Thm:DoFKUsers} for the 3-user Y-channel. Important insights about the optimal strategy for this channel can be obtained from a message-flow-graph (MFG), a graphical representation of the upper bounds which we introduce next.

\subsection{Message-flow-graph}
\label{Sec:MFG}
We first define the message-flow graph (MFG) formally, and then we discuss it in more detail.
\begin{definition}[Message-flow Graph]
To a $K$-user Y-channel and a desired DoF tuple $\d$ as defined in \eqref{Defpfvecd} corresponds a message-flow graph consisting of $K$ nodes and $K(K-1)$ directed edges, where the edge connecting nodes $i$ and $j\neq i$ has weight $d_{ij}$.
\end{definition}

\begin{figure}
\centering
\begin{tikzpicture}[semithick]
\node[circle] (c1) at (0,0) [fill=white, draw] {$1$};
\node[circle] (c2) at (3,0) [fill=white, draw] {$2$};
\node[circle] (c3) at (6,0) [fill=white, draw] {$3$};
\draw[->] (c1.east) to [out=20,in=160] node[above] {$2$} (c2.west) ;
\draw[->] (c2.west) to [out=200,in=-20] node[below] {$1$} (c1.east) ;
\draw[->] (c2.east) to node[below] {$1$}  (c3.west);
\draw[->] ($(c3.west)+(0,.2)$) to [out=135,in=45] node[above] {$1$}  ($(c1.east)+(0,.2)$);
\end{tikzpicture}
\caption{Message-flow-graph for a 3-user Y-channel with a DoF tuple $(2,0,1,1,1,0)$.}
\label{Fig:MessageFlow3UsersDHat}
\end{figure}
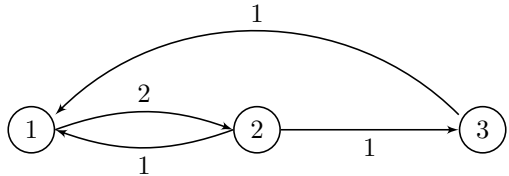

Now we describe the MFG with more detail. The message exchange in the Y-channel can be represented by an MFG as shown in Figure \ref{Fig:MessageFlow3UsersDHat}. In this graph, each node represents a user, and each directed-edge represents a message and is marked by the corresponding DoF (weight). Edges with zero weight are omitted for clarity.

\begin{remark}
Consider an expression as \eqref{Eq:DoFBoundThm}. The DoF components involved in this expression can be represented in an MFG as well. The MFG of this expression is similar to the one of the corresponding Y-channel, where an edge from node $i$ to node $j$ exists if $d_{ij}$ appears in the expression and has weight $d_{ij}$, and does not exist otherwise.
\end{remark}

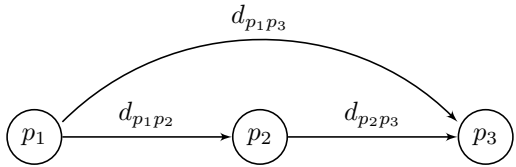
\begin{figure}
\centering
\begin{tikzpicture}[semithick]
\node[circle] (c1) at (0,0) [fill=white, draw] {$p_1$};
\node[circle] (c2) at (3,0) [fill=white, draw] {$p_2$};
\node[circle] (c3) at (6,0) [fill=white, draw] {$p_3$};
\draw[->] (c1.east) to node[above] {$d_{p_1p_2}$} (c2.west) ;
\draw[->] (c2.east) to node[above] {$d_{p_2p_3}$}  (c3.west);
\draw[->] ($(c1.east)+(0,.2)$) to [out=45,in=135] node[above] {$d_{p_1p_3}$}  ($(c3.west)+(0,.2)$);
\end{tikzpicture}
\caption{Message flow graph for the 3-user Y-channel representing the DoF upper bound \eqref{eq:DoF3User} leading to $d_{p_1p_2}+d_{p_2p_3}+d_{p_1p_3}\leq N$.}
\label{Fig:MessageFlow3Users}
\end{figure}

Consider for instance a 3-user Y-channel. According to Theorem \ref{Thm:DoFKUsers}, the DoF region of a 3-user MIMO Y-channel with $N\leq M$, denoted $\mathcal{D}_3$, is described by the following set of inequalities
\begin{align}
\label{eq:DoF3User}
\sum_{i=1}^{2}\sum_{j=i+1}^{3}d_{p_ip_j}\leq N, \quad \forall \p,
\end{align}
where $\p$ is a permutation of $(1,2,3)$. The upper bound \eqref{eq:DoF3User} bounds the DoF of the message exchange from $p_1$ to $p_2$, $p_1$ to $p_3$, and $p_2$ to $p_3$. This message exchange can be visualized using the MFG shown in Figure \ref{Fig:MessageFlow3Users}. The left-hand-side of \eqref{eq:DoF3User} can be obtained by summing the weights of the edges. Notice the following interesting property of this MFG: \emph{This MFG has no cycles}. We call this property the \emph{no-cycle property.}

This property is clearly true for any permutation $\p$. For instance, consider a specific permutation $\hat{\p}=(1,2,3)$. For this case, the upper bound above can be written as
\begin{align}
d_{12}+d_{13}+d_{23}\leq N.
\end{align}
A cycle would exist if we have $d_{21}$ or $d_{31}$ instead of $d_{13}$ leading to the cycles $1\to 2\to 1$ and $1\to 2\to 3\to 1$, respectively. The bound \eqref{eq:DoF3User} does not allow such cycles.

Let us assume that the outer bound \eqref{eq:DoF3User} is tight as claimed by Theorem \ref{Thm:DoFKUsers}. Under this assumption, the important insight obtained from the no-cycle property is that a DoF region optimal strategy for the Y-channel should have DoF constraints which do not constitute cycles. In other words, a strategy that imposes additional constraints, such as $d_{12}+d_{21}+d_{23}+d_{31}$ e.g., can not achieve the outer bound \eqref{eq:DoF3User}.

Now let us apply this insight on a 3-user MIMO Y-channel with $M=N=3$. Assume that we would like to achieve the DoF tuple \begin{align}
\label{eq:dp}
\d'=(2,0,1,1,1,0)
\end{align}
over this Y-channel. According to Theorem \ref{Thm:DoFKUsers}, $\d'$ is indeed achievable since it belongs to $\mathcal{D}_3$. 
How can we achieve this DoF tuple? To answer this question, let us start by examining a uni-directional strategy over the Y-channel.

\subsection{A uni-directional strategy}
In a uni-directional strategy, the operation of the relay is similar to the operation of a decode-forward (DF) relay \cite{CoverElgamal,ShaqfehQahtaniAlnuweiri} in a point-to-point relay channel where the message flow is uni-directional. Namely, the relay decodes all signals in the uplink, and re-transmits the signals to the respective destinations in the downlink\footnote{An amplify-forward strategy can also be used to achieve the same performance \cite{ParkAlouiniParkKo, CaoChongHoJorswieck}.}. Using such a uni-directional strategy, each signal consumes one dimension of the signal-space at the relay (1 DoF/dimension). Assume that one would want to achieve ${\d}'$ by using this strategy. In this case, the achievability of ${\d}'$ would require
\begin{align}
\label{Eq:UniDirBound3User}
d_\Sigma=d_{12}+d_{13}+d_{21}+d_{23}+d_{31}+d_{32}\leq N.
\end{align}
In other words, the total DoF should not be greater than the number of signal-space dimensions at the relay. This bound is not satisfied in this example since $d_\Sigma=5>N$ \eqref{eq:dp}. Thus, such a uni-directional strategy is not able to achieve ${\d}'$. 

Now let us analyse the bound \eqref{Eq:UniDirBound3User} by using an MFG. The MFG corresponding to this DoF constraint is shown in Figure \ref{Fig:MessageFlow3UsersDHat}. One can easily see that this MFG violates the no-cycle property since it has the cycles $1\to2\to1$ and $1\to2\to3\to1$. To achieve ${\d}'$, we need to use strategies which resolve such cycles. Let us first deal with the cycle $1\to2\to1$.

\subsection{A bi-directional strategy}
\label{Sec:BiDirStra3User}
We need a strategy which resolves the length-2 cycle ($2$-cycle) $1\to2\to1$ in \eqref{Eq:UniDirBound3User}, thus replacing the terms $d_{12}+d_{21}$ by some terms which do not constitute a $2$-cycle in the corresponding MFG. This can be achieved by using a bi-directional strategy as in the TWRC \cite{NamChungLee, NarayananPravinSprintson, AvestimehrSezginTse} as follows. Each pair of users align the signals they want to exchange over one dimension at the relay. Let users 1 and 2 send the signals $u_{12}$ and $u_{21}$, respectively, such that they align along one dimension at the relay. Thus, the relay can compute a linear combination of these symbols\footnote{Computation is performed using the compute-forward framework of \cite{NazerGastpar}.} $L(u_{12},u_{21})$ and forward this to users 1 and 2 in the downlink over one dimension. Then, each user can decode the desired signal after subtracting his own self-interference. This operation requires 1 dimension to send 2 signals, and is thus more efficient than the uni-directional strategy which requires 1 dimension per signal. 

By exchanging these two symbols, we use $d_{21}=1$ dimensions. The residual DoF tuple to be achieved is then $\d^*=(d_{12}-d_{21},0,0,d_{23},d_{31},0)$. Assume that one would try to achieve $\d^*$ using the uni-directional strategy thus requiring $d_{12}-d_{21}+d_{23}+d_{31}$ more dimensions. The resulting combination of bi-directional and uni-directional strategies would require $d_{12}+d_{23}+d_{31}$ dimensions at the relay. Since the relay has $N$ dimensions, this combination is possible if
\begin{align}
\label{Eq:BiUniDirBound3User}
d_{12}+d_{23}+d_{31}\leq N.
\end{align}
But this is not true since $d_{12}+d_{23}+d_{31}=4>N$ \eqref{eq:dp}. Although the use of the bi-directional strategy has reduced the required dimensions from 5 (uni-directional) to 4 (uni-and bi-directional), the DoF tuple $\d'$ is still not achievable.

At this point, it is worth to emphasize the role of the bi-directional strategy in `resolving' $2$-cycles. By comparing \eqref{Eq:UniDirBound3User} and \eqref{Eq:BiUniDirBound3User}, we can see that the 2-cycle in the MFG of the former has been resolved in the latter. However, the MFG of \eqref{Eq:BiUniDirBound3User} violates the no-cycle property as it has the $3$-cycle (cycle of length 3) $1\to2\to3\to1$. For this reason, the combination of uni- and bi-directionals strategies does not achieve $\d'$. To overcome this, we need a strategy that resolves this 3-cycle as given next.

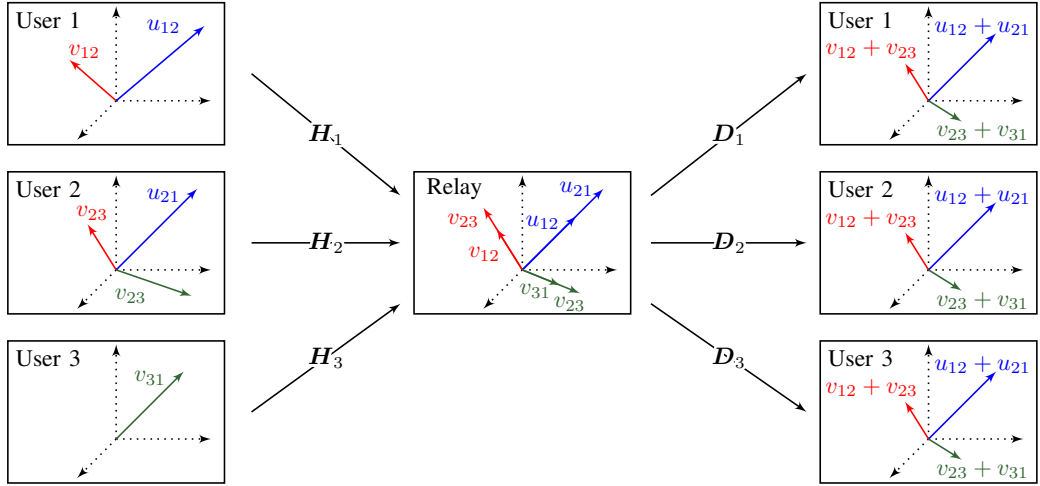
\begin{figure*}
\centering
\begin{tikzpicture}[semithick,scale=.9, every node/.style={scale=.9}]
\node (t1) at (0,0) {};
\node (t2) at (0,-2.5) {};
\node (t3) at (0,-5) {};
\node (r) at (6,-2.5) {};
\node (r1) at (12,0) {};
\node (r2) at (12,-2.5) {};
\node (r3) at (12,-5) {};
\node[fill=white] at ($(t1)+(-1.0,1.2,0)$) {User 1};
\node[fill=white] at ($(t2)+(-1.0,1.2,0)$) {User 2};
\node[fill=white] at ($(t3)+(-1,1.2,0)$) {User 3};
\node[fill=white] at ($(r)+(-1,1.2,0)$) {Relay};
\node[fill=white] at ($(r1)+(-1,1.2,0)$) {User 1};
\node[fill=white] at ($(r2)+(-1,1.2,0)$) {User 2};
\node[fill=white] at ($(r3)+(-1,1.2,0)$) {User 3};

\draw [] ($(t1)-(1.6,.65,0)$) rectangle ($(t1)+(1.6,1.45)$);
\draw [] ($(t2)-(1.6,.65,0)$) rectangle ($(t2)+(1.6,1.45)$);
\draw [] ($(t3)-(1.6,.65,0)$) rectangle ($(t3)+(1.6,1.45)$);
\draw [] ($(r)-(1.6,.65,0)$) rectangle ($(r)+(1.6,1.45)$);
\draw [] ($(r1)-(1.6,.65,0)$) rectangle ($(r1)+(1.6,1.45)$);
\draw [] ($(r2)-(1.6,.65,0)$) rectangle ($(r2)+(1.6,1.45)$);
\draw [] ($(r3)-(1.6,.65,0)$) rectangle ($(r3)+(1.6,1.45)$);

\foreach \i in {t1,t2,t3,r,r1,r2,r3}
{ 
\draw[->,dotted] (\i.center) to ($(\i)+(1.4,0,0)$);
\draw[->,dotted] (\i.center) to ($(\i)+(0,1.4,0)$);
\draw[->,dotted] (\i.center) to ($(\i)+(0,0,1.5)$);
}

\draw[->,blue] (t1.center) to ($(t1)+(1.5,1.3,.5)$);
\node[blue] at ($(t1)+(0.7,1.1,0)$) {$u_{12}$};
\draw[->,blue] (t2.center) to ($(t2)+(1.2,1.2,0)$);
\node[blue] at ($(t2)+(0.7,1.1,0)$) {$u_{21}$};
\draw[->,blue] (r.center) to ($(r)+(.8,0.8,0)$);
\node[blue] at ($(r)+(.3,0.7,0)$) {$u_{12}$};
\draw[->,blue] (r.center) to ($(r)+(1.2,1.2,0)$);
\node[blue] at ($(r)+(.8,1.2,0)$) {$u_{21}$};
\draw[->,blue] (r1.center) to ($(r1)+(1,1,0)$);
\node[blue] at ($(r1)+(.8,1.1,0)$) {$u_{12}+u_{21}$};
\draw[->,blue] (r2.center) to ($(r2)+(1,1,0)$);
\node[blue] at ($(r2)+(.8,1.1,0)$) {$u_{12}+u_{21}$};
\draw[->,blue] (r3.center) to ($(r3)+(1,1,0)$);
\node[blue] at ($(r3)+(.8,1.1,0)$) {$u_{12}+u_{21}$};

\draw[->,red] (t1.center) to ($(t1)+(0,1.3,1.8)$);
\node[red] at ($(t1)+(.3,1.5,2)$) {$v_{12}$};
\draw[->,red] (t2.center) to ($(t2)+(0,1.1,1.1)$);
\node[red] at ($(t2)+(.3,1.5,1.7)$) {$v_{23}$};
\draw[->,red] (r.center) to ($(r)+(0,1.0,1.0)$);
\node[red] at ($(r)+(-.1,.7,1.2)$) {$v_{12}$};
\draw[->,red] (r.center) to ($(r)+(0,1.5,1.5)$);
\node[red] at ($(r)+(-.1,1.5,2)$) {$v_{23}$};
\draw[->,red] (r1.center) to ($(r1)+(0,.9,.9)$);
\node[red] at ($(r1)+(-.3,1.3,1.4)$) {$v_{12}+v_{23}$};
\draw[->,red] (r2.center) to ($(r2)+(0,.9,.9)$);
\node[red] at ($(r2)+(-.3,1.3,1.4)$) {$v_{12}+v_{23}$};
\draw[->,red] (r3.center) to ($(r3)+(0,.9,.9)$);
\node[red] at ($(r3)+(-.3,1.3,1.4)$) {$v_{12}+v_{23}$};

\draw[->,Ogreen] (t2.center) to ($(t2)+(1.5,0,1.0)$);
\node[Ogreen] at ($(t2)+(0.6,0,1.0)$) {$v_{23}$};
\draw[->,Ogreen] (t3.center) to ($(t3)+(1,1,0)$);
\node[Ogreen] at ($(t3)+(.5,.9,0)$) {$v_{31}$};
\draw[->,Ogreen] (r.center) to ($(r)+(.8,0,0.6)$);
\node[Ogreen] at ($(r)+(0.5,0,.8)$) {$v_{31}$};
\draw[->,Ogreen] (r.center) to ($(r)+(1.2,0,0.9)$);
\node[Ogreen] at ($(r)+(1.2,0,1.3)$) {$v_{23}$};
\draw[->,Ogreen] (r1.center) to ($(r1)+(.8,0,.8)$);
\node[Ogreen] at ($(r1)+(1.3,0.05,1.3)$) {$v_{23}+v_{31}$};
\draw[->,Ogreen] (r2.center) to ($(r2)+(.8,0,.8)$);
\node[Ogreen] at ($(r2)+(1.3,0.05,1.3)$) {$v_{23}+v_{31}$};
\draw[->,Ogreen] (r3.center) to ($(r3)+(.8,0,.8)$);
\node[Ogreen] at ($(r3)+(1.3,0.05,1.3)$) {$v_{23}+v_{31}$};

\draw[->] ($(t1)+(2.0,0.4,0)$) to node[] {\contour{white}{$\H_{1}$}} ($(r)-(1.8,-1.1,0)$) ;
\draw[->] ($(t2)+(2.0,0.4,0)$) to node[] {\contour{white}{$\H_{2}$}} ($(r)-(1.8,-0.4,0)$) ;
\draw[->] ($(t3)+(2.0,.4,0)$) to node[] {\contour{white}{$\H_{3}$}} ($(r)-(1.8,0.5,0)$) ;
\draw[->] ($(r)+(1.9,1.1,0)$) to node[] {\contour{white}{$\D_{1}$}} ($(r1)-(1.8,-0.4,0)$) ;
\draw[->] ($(r)+(1.9,.4,0)$) to node[] {\contour{white}{$\D_{2}$}} ($(r2)-(1.8,-0.4,0)$) ;
\draw[->] ($(r)+(1.9,-.5,0)$) to node[] {\contour{white}{$\D_{3}$}} ($(r3)-(1.8,-0.4,0)$) ;

\end{tikzpicture}
\caption{A graphical illustration of the transmitter, relay, and receiver signal-space for the toy-example in Section \ref{Sec:3UserYChannel}. The relay computes the sum of the symbols received along each of the three directions, and forwards these sums. Each user is able to extract his desired signals after subtracting self-interference.}
\label{Fig:SigSpace3User}
\end{figure*}

\subsection{A cyclic strategy}
After assigning one dimension for bi-directional communication between users 1 and 2, two dimensions remain at the relay, and it remains to achieve $\d^*=(d_{12}-d_{21},0,0,d_{23},d_{31},0)=(1,0,0,1,1,0)$. In this case, users 1, 2, and 3 want to send a symbol each to users 2, 3, and 1, respectively. Denote these symbols by $v_{12}$, $v_{23}$, and $v_{31}$, respectively. Let users 1 and 2 send signals $v_{12}$ and $v_{23}$ such that they align along one dimension at the relay, and let users 2 and 3 send signals $v_{23}$ and $v_{31}$ such that they align along another dimension at the relay. Here, $v_{23}$ is sent twice by user 2, each time along a different direction. After reception, the relay computes linear combinations of these symbols $L_1(v_{12},v_{23})$ and 
$L_2(v_{23},v_{31})$, and then sends these linear combinations to the users in the downlink over two dimensions. After these combinations are received, user 1 decodes $v_{23}$ from $L_1$ after subtracting self-interference, and then decodes $v_{31}$ from $L_2$ after subtracting $v_{23}$. Similarly, users 2 and 3 obtain their desired signals. 

\begin{remark}
Note that this cyclic strategy is similar to the functional-decode-forward strategy in \cite{OngKellettJohnson_IT}, except that in our case, we perform the alignment over spatial sub-channels contrary to \cite{OngKellettJohnson_IT} which uses temporal sub-channels.
\end{remark}

This strategy requires only $d_{23}+d_{31}=2$ dimensions at the relay, contrary to the uni-directional strategy which requires 3 dimensions at the relay to deliver the same signals. The total number of required dimensions by the combination of the bi-directional and cyclic strategies is $d_{21}+d_{23}+d_{31}$. This should satisfy
\begin{align}
\label{Eq:BiCyBound3User}
d_{21}+d_{23}+d_{31}\leq N,
\end{align}
since the relay has $N$ dimensions in total. This constraint is satisfied by $\d'$ \eqref{eq:dp}. Thus, after this step, the DoF tuple $\d'$ is achieved. The resulting user and relay signal-space is as shown in Figure \ref{Fig:SigSpace3User}.

Now we can see the role of the cyclic strategy in resolving $3$-cycles. By comparing \eqref{Eq:BiUniDirBound3User} and \eqref{Eq:BiCyBound3User}, it is easy to see that the $3$-cycle in the MFG of the former has been resolved in the latter. The MFG of \eqref{Eq:BiCyBound3User} satisfies the no-cycle property, which was the desired goal in the first place. In conclusion, by designing a transmission strategy whose achievability is constrained by a DoF constraint which satisfies the no-cycle property, we could achieve the desired $\d'$. Although the uni-directional strategy was not needed in this particular example, in general, the optimal transmission strategy for the 3-user Y-channel is a combination of the three strategies (uni-directional, bi-directional, and cyclic).

It is due here to make the following note about the ordering of the strategies. It is important to start by allocating the DoF for the bi-directional strategy first, followed by the cyclic, and finally the uni-directional one. This follows from the ordering of the strategies in decreasing order of efficiency:
\begin{enumerate}
\item the bi-directional strategy consumes one dimension at the relay per two signals, for an efficiency of 2 DoF/dimension,
\item the cyclic strategy consumes two dimension at the relay per three signals, for an efficiency of 3/2 DoF/dimension,
\item the uni-directional strategy consumes one dimension at the relay per signals, for an efficiency of 1 DoF/dimension.
\end{enumerate}
This order will be used in the next section to prove the achievability of Theorem \ref{Thm:DoFKUsers} for the $K$-user case. Next, we extend this idea to the $K$-user Y-channel.

\section{Achievability of Theorem \ref{Thm:DoFKUsers}}
\label{Sec:Achievability}
In this section, we propose a transmission strategy which achieves the DoF region given in Theorem \ref{Thm:DoFKUsers}. The main components of the transmission strategy are channel diagonalization and a combination of bi-directional, cyclic, and uni-directional transmission strategies. The optimality of the given strategy is proved by proposing an optimal resource allocation based on the idea discussed in Section \ref{Sec:3UserYChannel} which we will extend to the $K$-user case. We start by describing channel diagonalization.

\subsection{Channel diagonalization}
\label{Sec:ChDiag}
Channel diagonalization is performed by using zero-forcing beam-forming with the aid of the Moore-Penrose pseudo inverse (MPPI). We need pre-coders that diagonalize all uplink channels, and also post-coders that diagonalize all downlink channels. 

Thus, the transmit signal of user $i$ is constructed as
\begin{align}
\x_i=\V_i\u_i,
\end{align}
where $\u_i\in\mathbb{C}^{N\times 1}$ is a vector which contains the codeword symbols satisfying $\tr(\mathbb{E}[\u_i\u_i^{H}])=\rho$, and where $\V_i\in\mathbb{C}^{M\times N}$ is the normalized right-MPPI of $\H_i$ given by
\begin{align}
\V_i=\alpha_i\H_i^\dagger,
\end{align}
with $\H_i^\dagger=\H_i^{H}[\H_i\H_i^{H}]^{-1}$ which exists if $N\leq M$, and with $\alpha_i=\|\H_i^{\dagger}\|_F^{-1}$ where $\|\H_i^\dagger\|_F$ is the Frobenius norm of $\H_i^\dagger$. This guarantees that $\x_i$ also satisfies the power constraint $\rho$, and that $\H_i\x_i=\H_i\V_i\u_i=\alpha_i\I_N\u_i$ thus achieving channel diagonalization in the uplink. The received signal at the relay is then
\begin{align}
\y_r=\sum_{i=1}^K\alpha_i\I_N\u_i+\z_r,
\end{align}
and over the $s$-th sub-channel, the relay receives
\begin{align}
y_{r,s}=\sum_{i=1}^K\alpha_iu_{i,s}+z_{r,s},
\end{align}
where $y_{r,s}$, $u_{i,s}$, and $z_{r,s}$ are the $s$-th components of $\y_r$, $\u_i$, and $\z_r$, respectively.

\begin{figure*}[t]
\centering
\subfigure[Uplink.]{
\begin{tikzpicture}[semithick]
\node (y1) at (-.5,4) [] {$u_{1,s}$};
\node (y2) at  (-.5,2) [] {$u_{2,s}$};
\node[draw] (r) at (3,2) [circle, minimum width=.5cm, minimum height=.5cm] {};
\node at (r) [rotate=0] {\large $+$};
\draw[->] (r.east) to ($(r.east)+(.5,0)$);
\draw[<-] (r.south) to ($(r.south)+(0,-.5)$);
\node at ($(r.east)+(.8,0)$) {$y_{r,s}$};
\node at ($(r.south)+(0,-.8)$) {$z_{r,s}$};
\node (y3) at (-.5,0) [] {$u_{3,s}$};
\draw[<-] ($(r.west)+(0,.25)$) to node[] {\contour{white}{$\alpha_1$}} (y1.east);
\draw[<-] ($(r.west)+(0,0)$) to node[] {\contour{white}{$\alpha_2$}} (y2.east);
\draw[<-] ($(r.west)+(0,-.25)$) to node[] {\contour{white}{$\alpha_3$}} (y3.east);
\node[] at (1.8,1.55) [rectangle, minimum width=5.5cm, minimum height=5.7cm] {};
\end{tikzpicture}
}
\hspace{1cm}
\subfigure[Downlink.]{
\begin{tikzpicture}[semithick]
\node[draw] (y1) at (3,4) [circle, minimum width=.5cm, minimum height=.5cm] {};
\node at (y1) [rotate=0] {\large $+$};
\draw[->] (y1.east) to ($(y1.east)+(.5,0)$);
\draw[<-] (y1.south) to ($(y1.south)+(0,-.5)$);
\node at ($(y1.east)+(.8,0)$) {$\tilde{y}_{1,s}$};
\node at ($(y1.south)+(0,-.8)$) {$\tilde{z}_{1,s}$};
\node[draw] (y2) at (3,2) [circle, minimum width=.5cm, minimum height=.5cm] {};
\node at (y2) [rotate=0] {\large $+$};
\draw[->] (y2.east) to ($(y2.east)+(.5,0)$);
\draw[<-] (y2.south) to ($(y2.south)+(0,-.5)$);
\node at ($(y2.east)+(.8,0)$) {$\tilde{y}_{2,s}$};
\node at ($(y2.south)+(0,-.8)$) {$\tilde{z}_{2,s}$};
\node[] (r) at (-.5,2) {$x_{r,s}$};
\node[draw] (y3) at (3,0) [circle, minimum width=.5cm, minimum height=.5cm] {};
\node at (y3) [rotate=0] {\large $+$};
\draw[->] (y3.east) to ($(y3.east)+(.5,0)$);
\draw[<-] (y3.south) to ($(y3.south)+(0,-.5)$);
\node at ($(y3.east)+(.8,0)$) {$\tilde{y}_{3,s}$};
\node at ($(y3.south)+(0,-.8)$) {$\tilde{z}_{3,s}$};
\draw[->] ($(r.east)+(0,.25)$) to (y1.west) ;
\draw[->] ($(r.east)+(0,0)$) to (y2.west) ;
\draw[->] ($(r.east)+(0,-.25)$) to (y3.west);
\node[] at (1.8,1.55) [rectangle, minimum width=5.5cm, minimum height=5.7cm] {};
\end{tikzpicture}
}
\caption{A 3-user MIMO Y-channel after pre- and post-processing using the MPPI. The channels matrices are diagonalized, and thus decomposed into $N$ parallel SISO Y-channels. The figure shows the $s$-th sub-channel.}
\label{Fig:Diagonalization}
\end{figure*}
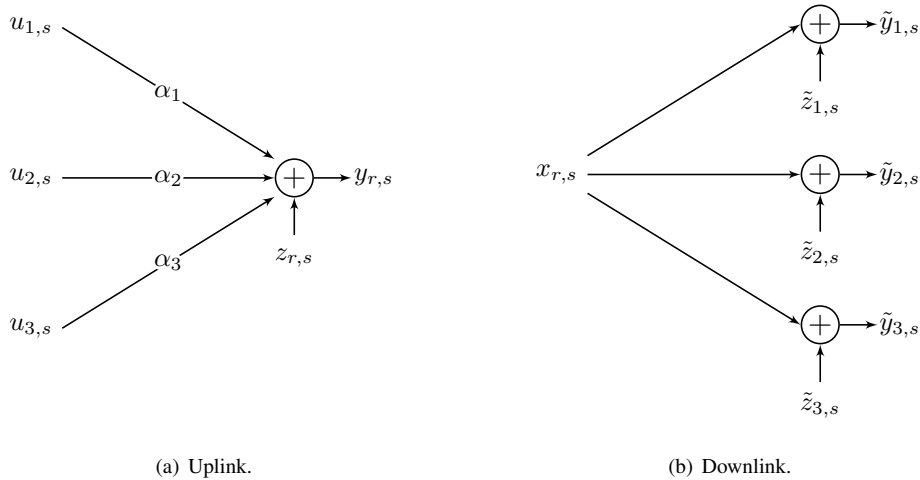

In the downlink, the users use a post-coding matrix $\U_i\in\mathbb{C}^{N\times M}$ given by the left-MPPI of $\D_i$, i.e.,
\begin{align}
\U_i=[\D_i^{H}\D_i]^{-1}\D_i^{H},
\end{align}
which exists if $N\leq M$. The processed received signal at user $i$ is thus
\begin{align}
\tilde{\y}_i=\U_i\y_i=\I_N\x_r+\U_i\z_i=\I_N\x_r+\tilde{\z}_i
\end{align}
which achieves channel diagonalization in the downlink. Over the $s$-th sub-channel, the user receives
\begin{align}
\tilde{y}_{i,s}=x_{r,s}+\tilde{z}_{j,s},
\end{align}
where $\tilde{y}_{i,s}$, $x_{r,s}$, and $\tilde{z}_{i,s}$ are the $s$-th components of $\tilde{\y}_{i}$, $\x_r$, and $\tilde{\z}_{i}$, respectively. Note that the noise $\tilde{\z}_{i}$ is colored in general, since $\mathbb{E}[\tilde{\z}_{i}\tilde{\z}_{i}^H]$ is not a diagonal matrix. Although this noise correlation can be exploited at the receiver to increase the achievable rate, this is not necessary from a DoF point of view. Thus, we can assume that these noises are independent, which delivers a worst-case performance.

The result of this diagonalization is a decomposition of the MIMO Y-channel into $N$ parallel SISO Y-channels as shown in Figure \ref{Fig:Diagonalization}. From this point on, we deal with the MIMO Y-channel after pre- and post-coding as a set of $N$ parallel SISO Y-channels. Now let us describe the transmission strategies to be used over these sub-channels.

\subsection{Transmission strategies}
\label{Sec:TranStra}
In this subsection, we describe the different communication strategies that will be used to achieve the DoF region of the Y-channel. Cycles will play an important role in the discussion in this subsection and the next one. So we start by introducing some notation related to cycles.

\subsubsection{Cycle notation}
An $\ell$-cycle $i_1\to i_2\to\cdots\to i_\ell\to i_1$ is denoted by the tuple $\c_\ell=(i_1,i_2,\cdots,i_\ell)$. Note that this notation is cyclic-shift invariant. In other words, if $\phi_\eta(\c_\ell)$ is a cyclic-shift of $\c_\ell$ by $\eta$ positions, then $\c_\ell$ and $\phi_\eta(\c_\ell)$ are equivalent cycles for all $\eta=1,\cdots,\ell-1$. Let us denote the set of all distinct $\ell$-cycles in the $K$-user Y-channel by $\mathcal{S}_\ell$. This set contains all $\ell$-tuples which are not cyclically equivalent, i.e.,
\begin{align}
\label{eq:Sl}
\c_\ell,\hat{\c}_\ell\in\mathcal{S}_\ell\Rightarrow \c_\ell,\hat{\c}_\ell\in\mathcal{K}^\ell \text{ and } \c_\ell\neq \phi_\eta(\hat{\c}_\ell)\ \forall \eta=1,\cdots,\ell-1.
\end{align}
Recall that $\mathcal{K}=\{1,\cdots,K\}$. The cardinality of $\mathcal{S}_\ell$ is given by $|\mathcal{S}_\ell|=\frac{K!}{\ell\cdot(K-\ell)!}$, which is the number of permutations with $\ell$ elements from $\mathcal{K}$ given by $\frac{K!}{(K-\ell)!}$ divided by the number of cyclically equivalent permutations $\ell$. We denote the $n$-th element of $\mathcal{S}_\ell$ by $\c_{\ell[n]}$.

We also denote by $\mathcal{E}_{\c_{\ell[n]}}$ the set of all edges of the cycle $\c_{\ell [n]}$, i.e., for $\c_{\ell [n]}\in\mathcal{S}_\ell $,
\begin{align}
\label{eq:Ecln}
\mathcal{E}_{\c_{\ell [n]}}=\{\c_{\ell [n],1}\c_{\ell [n],2},\ \c_{\ell [n],2}\c_{\ell [n],3},\cdots,\c_{\ell [n],\ell -1}\c_{\ell [n],\ell},\ \c_{\ell [n],\ell}\c_{\ell [n],1}\},
\end{align}
where $\c_{\ell [n],b}$ is the $b$-th component of $\c_{\ell[n]}$. Note that we denote the edges by $\c_{\ell [n],b}\c_{\ell [n],b+1}$ instead of the more common $(\c_{\ell [n],b},\c_{\ell [n],b+1})$ in order to avoid confusion with the 2-cycle $(\c_{\ell [n],b},\c_{\ell [n],b+1})$. For instance, the set of edges of the cycle $\c_{3[1]}=(1,2,3)$ is given by $\mathcal{E}_{\c_{3[1]}}=\{12,\ 23,\ 31\}$. We also denote by $\mathcal{E}$ the set of all possible edges of the MFG of the Y-channel. This set can be written as
\begin{align}
\label{Eq:DefE}
\mathcal{E}=\bigcup_{n=1}^{|\mathcal{S}_2|} \mathcal{E}_{\c_{2[n]}},
\end{align}
since the union of the sets of edges of all $2$-cycles covers all the edges of the MFG. 

%
%

The rest of this subsection is split into three parts. We start be describing the bi-directional strategy (or $2$-cyclic strategy referring to communication over $2$-cycles), then we describe the $\ell$-cyclic strategy (communication over $\ell$-cycles, $\ell>2$), and finally, we describe the uni-directional strategy.

\subsubsection{Bi-directional strategy ($2$-cyclic)}
\label{Sec:BiDirStrategy}
In the bi-directional strategy, communication over each sub-channel is similar to communication over the SISO TWRC \cite{NamChungLee_IT}. Consider the $2$-cycle $\c_{2[n]}=(i,j)$, $n\in\{1,\cdots,|\mathcal{S}_2|\}$ e.g., where the communicating partners want to exchange one symbol with each other. For this cycle, users $i$ and $j$ send symbols $u_{i,\c_{2[n]}},u_{j,\c_{2[n]}}\in\mathbb{C}$, respectively, to the relay over the $s$-th sub-channel in the uplink. These users set $u_{i,s}=u_{i,\c_{2[n]}}$ and $u_{j,s}=u_{j,\c_{2[n]}}$. The remaining users do not send over this sub-channel. The relay receives 
\begin{align*}
y_{r,s}=\alpha_iu_{i,\c_{2[n]}}+\alpha_ju_{j,\c_{2[n]}}+z_{r,s}.
\end{align*} 
After receiving $\tau$ instances of this signal, i.e., $y_{r,s}^\tau$ where $\tau$ is the code length, the relay computes\footnote{Computation at the relay can be enabled by encoding the signals using nested-lattice codes as in \cite{ChaabanSezgin_IT_IRC}. As the discussion on lattice codes is not within the scope of the paper, the interested reader is referred to \cite{NazerGastpar}. From a DoF perspective, a similar performance can be achieved by using quantize-forward or compress-forward \cite{AvestimehrSezginTse, GunduzTuncelNayak}, and also by using amplify-forward \cite{VazeHeath}.} $\alpha_iu_{i,\c_{2[n]}}^\tau+\alpha_ju_{j,\c_{2[n]}}^\tau$ (see Appendix \ref{App:Rate2DoF}). The relay then forwards this sum to the two users over sub-channel $s$ in $\tau$ channel uses of the downlink after multiplying by a normalization factor $\gamma_s$ for power allocation. Thus, the relay sets $\x_{r,s}=\gamma_s(\alpha_iu_{i,\c_{2[n]}}+\alpha_ju_{j,\c_{2[n]}})$. User $i$ receives
\begin{align*}
\tilde{y}_{i,s}=\gamma_s(\alpha_iu_{i,\c_{2[n]}}+\alpha_ju_{j,\c_{2[n]}})+\tilde{z}_{i,s},
\end{align*}
from which the desired signal is decoded after self-interference cancellation. User $j$ obtains his desired signals similarly. Since each user can decode his desired signal reliably, this guarantees the achievability of 1 DoF per user (users $i$ and $j$) over one sub-channel (see Appendix \ref{App:Rate2DoF} for more details). If these users would like to achieve $d$ DoF (each) in this transmission, then a bundle of $d$ sub-channels is used both in the uplink and in the downlink to exchange a total of $2d$ symbols. The efficiency of this strategy is thus 2 DoF/dimension.

\subsubsection{$\ell$-cyclic strategy}
\label{Sec:kCycStrategy}
Consider the $\ell$-cycle $\c_\ell=(i_1,i_2,\cdots,i_\ell)$, where each user wants to send one symbol to the next user in the cycle (with cyclic indexing). In the $\ell$-cyclic strategy, users $i_q$ and $i_{q+1}$ send the symbols $u_{i_q,\c_\ell}$ and $u_{i_{q+1},\c_\ell}$ intended to user $i_{q+1}$ and $i_{q+2}$, respectively, over one sub-channel $s_q$ in the uplink with $q=1,\cdots,\ell-1$, by setting $u_{i_q,s_q}=u_{i_q,\c_\ell}$ and $u_{i_{q+1},s_{q+1}}=u_{i_{q+1},\c_\ell}$. The symbol $u_{i_\ell,\c_\ell}$ is intended to user $i_1$ (cyclic flow). Note that using this strategy, users $i_2\cdots,i_{\ell-1}$ repeat their symbols twice over two sub-channels, leading to dependent coding over sub-channels. The relay receives the following signal
\begin{align*}
y_{r,s_q}=\alpha_{i_q}u_{i_q,\c_\ell}+\alpha_{i_{q+1}}u_{i_{q+1},\c_\ell}+z_{r,s_q}
\end{align*}
over sub-channel $s_q$. It computes the sum $\alpha_{i_q}u_{i_q,\c_\ell}+\alpha_{i_{q+1}}u_{i_{q+1},\c_\ell}$ for all $q$. Then it normalizes this sum by $\gamma_{s_q}$ to fulfil the power constraint, and sends it over the sub-channel $s_q$ in the downlink. User $i_p$, $p=1,\cdots,\ell$, receives 
\begin{align*}
\tilde{y}_{i_p,s_q}
=\gamma_{s_q}(\alpha_{i_q}u_{i_q,\c_\ell}+\alpha_{i_{q+1}}u_{i_{q+1},\c_\ell})+\tilde{z}_{i_p,s_q},
\end{align*} 
for all $q=1,\cdots,\ell-1$. Each user can extract all signals exchanged by the $\ell$-cyclic strategy. User $i_p$ starts by decoding $u_{i_{p+1},\c_\ell}$ from sub-channel $s_p$ after cancelling self-interference $u_{i_p,\c_\ell}$. Then it continues to sub-channel $s_{p+1}$ to decode $u_{i_{p+2},\c_\ell}$ after cancelling the already decoded $u_{i_{p+1},\c_\ell}$, and so on, until all symbols are decoded. Since all desired symbols can be decoded reliably, this guarantees the achievability of 1 DoF per user (cf. Appendix \ref{App:Rate2DoF}). Hence, a total of $\ell$ DoF over $\ell-1$ sub-channels is achieved. If each user wants to send $d$ streams to the next users in the cycle, then a bundle of $d$ sub-channels is used for each signal-pair in the uplink and in the downlink. In total this requires $(\ell-1)d$ sub-channels for exchanging $\ell d$ symbols. Thus, the efficiency of this $\ell$-cyclic strategy is $\ell/(\ell-1)$ DoF/dimension.

\begin{remark}
The bi-directional strategy can also be interpreted as an $\ell$-cyclic strategy with $\ell=2$.
\end{remark}

\subsubsection{Uni-directional strategy}
\label{Sec:StraUniDir}
The uni-directional strategy is a simple decode-forward strategy (or amplify-forward strategy \cite{ParkAlouiniParkKo, CaoChongHoJorswieck}). In this strategy, each user sends $d$ symbols to the desired destination over non-shared
$d$ sub-channels in the uplink and $d$ sub-channels in the downlink. The efficiency of this strategy is thus 1 DoF/dimension.

\begin{table}
\centering
\begin{tabular}{| c || c | c |c |}
    \hline
    Transmission & dimensions  & symbols  & efficiency \\
     strategy &  required &  delivered &  (symbols/dimension)\\\hline
    bi-directional & $1$ & $2$ & $2$\\\hline
    $3$-cyclic & $2$ & $3$ & $3/2$\\\hline
    $4$-cyclic & $3$ & $4$ & $4/3$\\\hline
    $\vdots$ & $\vdots$ & $\vdots$ & $\vdots$\\\hline
    $K$-cyclic & $K-1$ & $K$ & $K/(K-1)$\\\hline
    Uni-directional & $1$ & $1$ & $1$\\\hline
    \end{tabular}
    \caption{The schemes required to achieve the DoF region of the $K$-user MIMO Y-channel with $N\leq M$ listed in decreasing order of efficiency.}
    \label{Tab:Schemes}
\end{table}

These strategies are collected in Table \ref{Tab:Schemes} in decreasing order of efficiency. The next goal is to allocate signals appropriately over the $N$ sub-channels of the Y-channel in a way that achieves any DoF tuple in the DoF region $\mathcal{D}_K$ described by
\begin{align}
\label{eq:DoFRegionKUser}
\sum_{i=1}^{K-1}\sum_{j=i+1}^{K}d_{p_ip_j}\leq N, \quad \forall \mathbf{p}
\end{align}
where $\mathbf{p}$ is a permutation of $(1,\cdots,K)$ and $p_i$ is its $i$-th component as given in Theorem \ref{Thm:DoFKUsers}. This problem can be interpreted as a resource allocation problem where the available resources are the $N$ dimensions provided by the $N$ sub-channels. An optimal resource allocation strategy is provided in the next subsection.

\subsection{Resource allocation}
\label{Sec:ResAlloc}
After channel diagonalization, the problem of the DoF region achievability reduces to a resource allocation problem. We have $N$ dimensions as resources, which need to be shared by the users in an optimal way. The resource allocation is performed similar to the 3-user example in Section \ref{Sec:3UserYChannel}. Here, we discuss the $K$-user case.

For a $K$-user MIMO Y-channel with $N\leq M$, we need to show the achievability of any DoF tuple $\d$ which satisfies \eqref{eq:DoFRegionKUser}. Recall that the DoF region $\mathcal{D}_K$ is described by DoF upper bounds that do not constitute any cycles (Section \ref{Sec:MFG}). On the other hand, a DoF tuple $\d\in\mathcal{D}_K$ might constitute cycles (cf. Figure \ref{Fig:MessageFlow3UsersDHat}). As described in Section \ref{Sec:3UserYChannel}, the achievability of all DoF tuples in $\mathcal{D}_K$ requires strategies that resolve such cycles. In the $K$-user Y-channel, we have cycles of length $2$ up to $K$. Next, we describe how these cycles can be resolved, and we prove the achievability of any $\d\in\mathcal{D}_K$. A pseudo-code which describes the achievability of $ḑ\in\mathcal{D}_K$ is given in Algorithm \ref{Alg}. Since the bi-directional strategy is the most efficient among the set of strategies in Table \ref{Tab:Schemes}, we start by allocating resources to this strategy first.

\RestyleAlgo{boxruled}
\begin{algorithm}[t]
\SetKwInOut{Input}{input}\SetKwInOut{Output}{output}
\Input{$\d\in\mathcal{D}_K$}
\For{$\ell\leftarrow 2$ \KwTo $K$}{
Generate $\mathcal{S}_\ell=\{\c_{\ell[1]},\c_{\ell[2]},\cdots,\c_{\ell[|\mathcal{S}_\ell|]}\}$ according to \eqref{eq:Sl}\;
\For{$n\leftarrow 1$ \KwTo $|\mathcal{S}_\ell|$}{
Calculate $d_{\c_{\ell [n]}}$ according to \eqref{eq:dcln}\;
Apply $\ell$-cyclic strategy according to Sec. \ref{Sec:kCycStrategy}-\ref{Sec:BiDirStrategy}\;
}
}
Generate $\mathcal{E}$ according to \eqref{Eq:DefE}\;
\For{$\e\in\mathcal{E}$}{
Calculate $d^u_{\e}$ according to \eqref{eq:du}\;
Apply uni-directional strategy according to Sec. \ref{Sec:StraUniDir}\;
}
\caption{DoF-region achieving scheme}\label{Alg}
\end{algorithm}

\subsubsection{Resource allocation for the bi-directional strategy}
The bi-directional strategy will be used to resolve $2$-cycles. To this end, for each $2$-cycle $\c_{2[n]}$, $n=1,\cdots,|\mathcal{S}_2|$, we allocate the DoF to the bi-directional strategy according to
\begin{align}
\label{eq:DoF2Cycle}
d_{\c_{2[n]}}=\min_{\e\in\mathcal{E}_{\c_{2[n]}}}\left\{d_{\e}\right\},
\end{align}
where $d_{\e}$ represents component of $\d$ corresponding to edge $\e$. In other words, each user in a $2$-cycle achieves $d_{\c_{2[n]}}$ DoF by using the bi-directional strategy over $d_{\c_{2[n]}}$ sub-channels. Consider the cycle $\c_{2[1]}=(1,2)$ with edges $\mathcal{E}_{\c_{2[1]}}=\{12, 21\}$ for instance. For this $2$-cycle, we get $d_{(1,2)}=\min\{d_{12},d_{21}\}$, which determines the DoF to be achieved by each of users 1 and 2 using the bi-directional strategy. The involved partners in this cycle ($\c_{2[n],1}$ and $\c_{2[n],2}$) apply the bi-directional strategy over $d_{\c_{2[n]}}$ sub-channels as described in Section \ref{Sec:BiDirStrategy}.

\subsubsection{Resource allocation for the $3$-cyclic strategy}
After allocating resources to $2$-cycles, $K(K-1)/2$ components of the desired DoF tuple $\d$ are achieved. The residual DoF tuple to be achieved has at most $K(K-1)/2$ non-zero components. Namely, if users $i$ and $j$ want to exchange $d_{ij}$ and $d_{ji}\geq d_{ij}$ symbols for distinct $i,j\in\mathcal{K}$, after using the bi-directional strategy, $d_{ij}$ symbols from each of user $i$ and $j$ are successfully exchanged. However, $d_{ji}-d_{ij}\geq0$ symbols remain to be sent from user $j$ to $i$. Thus, $d_{ji}$ is only partially achieved. 

\begin{figure}
\centering
\subfigure[Initial MFG.]{
\begin{tikzpicture}[semithick, scale=.7]
\node[circle] (c1) at (0,0) [fill=white, draw] {$1$};
\node[circle] (c2) at (3,0) [fill=white, draw] {$2$};
\node[circle] (c3) at (6,0) [fill=white, draw] {$3$};
\node[circle] (c4) at (9,0) [fill=white, draw] {$4$};
\draw[->] (c1.east) to [out=20,in=160] node[above] {$3$} (c2.west);
\draw[->] (c2.west) to [out=200,in=-20] node[below] {$1$} (c1.east);
\draw[->] (c2.east) to [out=20,in=160] node[above] {$2$} (c3.west);
\draw[->] (c3.west) to [out=200,in=-20] node[below] {$1$} (c2.east);
\draw[->] ($(c3.west)+(0,.2)$) to [out=135,in=45] node[above] {$1$}  ($(c1.east)+(0,.2)$);
\draw[->] ($(c2.east)+(0,.2)$) to [out=45,in=135] node[above] {$1$}  ($(c4.west)+(0,.2)$);
\draw[->] ($(c4.west)+(0,0)$) to [out=-135,in=-45] node[above] {$2$}  ($(c1.east)-(0,.2)$);
\end{tikzpicture}
}
\hspace{.5cm}
\subfigure[MFG after taking care of $2$-cycles.]{
\begin{tikzpicture}[semithick, scale=.7]
\node[circle] (c1) at (0,0) [fill=white, draw] {$1$};
\node[circle] (c2) at (3,0) [fill=white, draw] {$2$};
\node[circle] (c3) at (6,0) [fill=white, draw] {$3$};
\node[circle] (c4) at (9,0) [fill=white, draw] {$4$};
\draw[->] (c1.east) to [out=20,in=160] node[above] {$2$} (c2.west);
\draw[->] (c2.east) to [out=20,in=160] node[above] {$1$} (c3.west);
\draw[->] ($(c3.west)+(0,.2)$) to [out=135,in=45] node[above] {$1$}  ($(c1.east)+(0,.2)$);
\draw[->] ($(c2.east)+(0,.2)$) to [out=45,in=135] node[above] {$1$}  ($(c4.west)+(0,.2)$);
\draw[->] ($(c4.west)+(0,0)$) to [out=-135,in=-45] node[above] {$2$}  ($(c1.east)-(0,.2)$);
\end{tikzpicture}
\label{Fig:SigFlow4User2}
}
\subfigure[MFG after taking care of $2$-cycles and the $3$-cycle $(1,2,3)$.]{
\begin{tikzpicture}[semithick, scale=.7]
\node[circle] (c1) at (0,0) [fill=white, draw] {$1$};
\node[circle] (c2) at (3,0) [fill=white, draw] {$2$};
\node[circle] (c3) at (6,0) [fill=white, draw] {$3$};
\node[circle] (c4) at (9,0) [fill=white, draw] {$4$};
\draw[->] (c1.east) to [out=20,in=160] node[above] {$1$} (c2.west);
\draw[->] ($(c2.east)+(0,.2)$) to [out=45,in=135] node[above] {$1$}  ($(c4.west)+(0,.2)$);
\draw[->] ($(c4.west)+(0,0)$) to [out=-135,in=-45] node[above] {$2$}  ($(c1.east)-(0,.2)$);
\end{tikzpicture}
\label{Fig:SigFlow4User2}
}
\caption{A MFG for a 4-user Y-channel representing a DoF tuple with $d_{12}=3$, $d_{23}=d_{41}=2$, and  $d_{21}=d_{24}=d_{31}=d_{32}=1$. The MFG before and after DoF allocation for $2$-cycles is shown. In Fig. \ref{Fig:SigFlow4User2}, we can see that the cycles $(1,2,3)$ and $(1,2,4)$ share the edge $12$. These two $3$-cycles are resolved by the $3$-cyclic strategy with $d_{(1,2,3)}=d_{(1,2,4)}=1$.}
\label{Fig:MessageFlow4Users}
\end{figure}
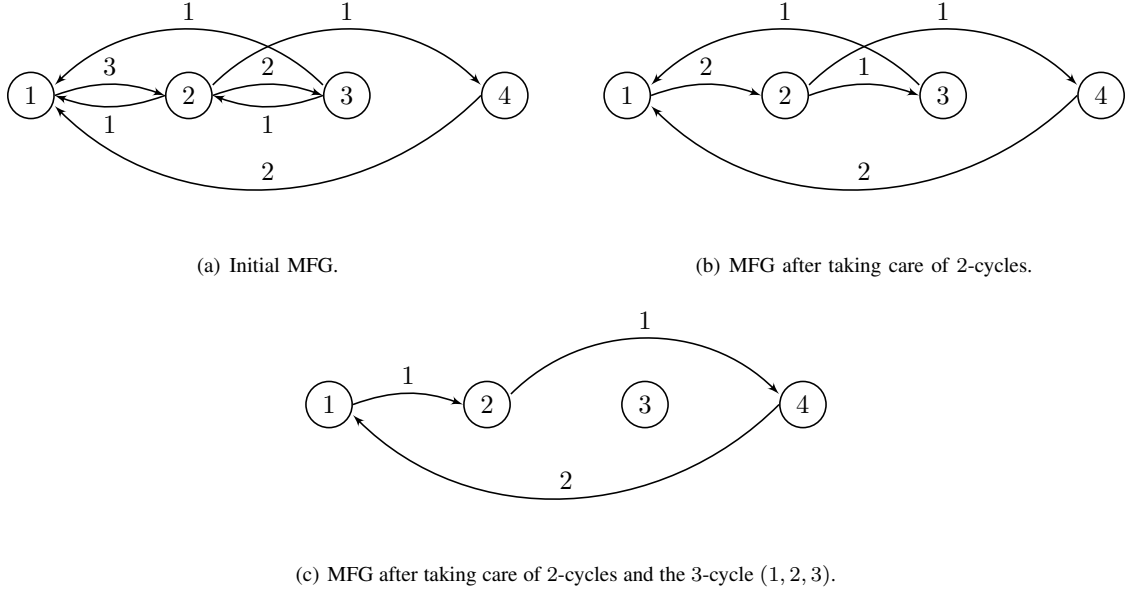

The residual DoF tuple might constitute cycles of length 3 or more. We resolve $3$-cycles since the $3$-cyclic strategy which is the second best strategy in terms of efficiency. Consider a $3$-cycle $\c_{3[n]}$, $n=1,\cdots,|\mathcal{S}_3|$. We allocate resources to the $3$-cyclic strategy corresponding to this $3$-cycle as follows
\begin{align}
\label{eq:dc3m}
d_{\c_{3[n]}}&=\min_{\e\in\mathcal{E}_{\c_{3[n]}}}
\left\{
d_{\e}
-\sum_{m=1}^{|\mathcal{S}_2|}\ind_{\mathcal{E}_{\c_{2[m]}}}(\e)d_{\c_{2[m]}}-\sum_{m=1}^{n-1}\ind_{\mathcal{E}_{\c_{3[m]}}}(\e)d_{\c_{3[m]}}
\right\}.
\end{align}
With this allocation, each user in the $3$-cycle $\c_{3[n]}$ achieves $d_{\c_{3[n]}}$ DoF, and the corresponding $3$-cyclic strategy is performed over $2d_{\c_{3[n]}}$ sub-channels. Here, $\ind_{\mathcal{E}_{\c_{2[m]}}}(\e)$ is an indicator function which is equal to 1 if $\e\in\mathcal{E}_{\c_{2[m]}}$ and 0 otherwise. The first sum in \eqref{eq:dc3m} represents the DoF allocated to $2$-cycles sharing the edge $\e$ with $\c_{3[n]}$, and the second one represents the DoF that have been already allocated to $3$-cycles $\c_{3[m]}$, $m<n$, sharing the edge $\e$ with $\c_{3[n]}$. As an example, assume that after allocating resources for $2$-cycles in a 4-user Y-channel, we end up with a residual DoF tuple with cycles $\c_{3[1]}=(1,2,3)$ and $\c_{3[2]}=(1,2,4)$ (see Figure \ref{Fig:MessageFlow4Users}). We subsequently set
\begin{align*}
d_{\c_{3[1]}}&=\min\{d_{12}-d_{(1,2)},d_{23}-d_{(2,3)},d_{31}-d_{(1,3)}\},\\
d_{\c_{3[2]}}&=\min\{d_{12}-d_{(1,2)}-d_{\c_{3[1]}},d_{24}-d_{(2,4)},d_{41}-d_{(1,4)}\},
\end{align*}
so that each user in the $3$-cycles $\c_{3[1]}$ and $\c_{3[2]}$ achieves $d_{\c_{3[1]}}$ and $d_{\c_{3[2]}}$ DoF by using the $3$-cyclic strategy over $2d_{\c_{3[1]}}$ and $2d_{\c_{3[2]}}$ sub-channels, respectively. This resolves all $3$-cycles in Figure \ref{Fig:MessageFlow4Users}.

After allocating resources to the $3$-cyclic strategy, we obtain the number of sub-channels to be used for each $3$-cycle. The transmission of the corresponding signals is done as described in Section \ref{Sec:kCycStrategy}. The cycles of length $4$ to $K$ can be treated similarly. Next, we illustrate the resource allocation for a general $\ell$-cycle strategy.

\subsubsection{Resource allocation for the $\ell$-cyclic strategy}
After handling all cycles of length $2$ to $\ell-1$, we consider $\ell$-cycles, $\ell=3,\cdots,K$. Consider an $\ell$-cycle $\c_{\ell [n]}$, $n=1,\cdots,|\mathcal{S}_\ell|$. We allocate the DoF to the $\ell$-cyclic strategy corresponding to this $\ell$-cycle as follows
\begin{align}
\label{eq:dcln}
d_{\c_{\ell [n]}}=\min_{\e\in\mathcal{E}_{\c_{\ell [n]}}}
\left\{
d_{\e}
-\sum_{i=2}^{\ell-1}
\sum_{m=1}^{|\mathcal{S}_i|}\ind_{\mathcal{E}_{\c_{i[m]}}}(\e)d_{\c_{i[m]}}-\sum_{m=1}^{n-1}\ind_{\mathcal{E}_{\c_{\ell[m]}}}(\e)d_{\c_{\ell [m]}}
\right\}.
\end{align}
Using this allocation, the users in the $\ell$-cycle $\c_{\ell [n]}$ achieve $d_{\c_{\ell [n]}}$ DoF each, by using the $\ell$-cyclic strategy over $(\ell-1)d_{\c_{\ell [n]}}$ sub-channels. In \eqref{eq:dcln}, we subtract from $d_{\e}$ all the DoF that have been allocated to $i$-cycles, $i=2,\cdots,\ell-1$, sharing edge $\e$ with $\c_{\ell [n]}$, and we also subtract the DoF that have been allocated to previous $\ell$-cycles ($\c_{\ell [m]}$, $m=1,\cdots,n-1$) sharing the edge $\e$ with $\c_{\ell [n]}$. The allocated DoF for the $\ell$-cyclic strategy are achieved as described in Section \ref{Sec:kCycStrategy}.

\subsubsection{Uni-directional strategy}
After considering all cycles of length $2$ to $K$, there might still remain some residual DoF tuple that need to be achieved. This is achieved using the uni-directional strategy. The remaining DoF to be achieved by the uni-directional strategy from user $i$ to user $j$ can be expressed as
\begin{align}
\label{eq:du}
d_{\e}^u=d_{\e}
-\sum_{\ell=2}^{K}
\sum_{m=1}^{|\mathcal{S}_\ell|}\ind_{\mathcal{E}_{\c_{\ell [m]}}}(ij)d_{\c_{\ell [m]}},
\end{align}
where $\e\in\mathcal{E}$ represents the edge $ij$. At this point, the description of the resource allocation is complete. Next, we show that this resource allocation is in fact optimal, and achieves any DoF tuple $\d$ in the DoF region $\mathcal{D}_K$ defined in Theorem \ref{Thm:DoFKUsers}.

\subsection{Optimality}
The question that remains is on the optimality of the resource allocation presented above. We show that this resource allocation is indeed optimal, and obtain the following lemma.
\begin{lemma}
\label{Lem:Opt}
The resource allocation strategy presented in Section \ref{Sec:ResAlloc} is optimal, i.e., achieves every $\d\in\mathcal{D}_K$.
\end{lemma}
To prove this, we have to show that the number of sub-channels suffices for all bi-directional, $\ell$-cyclic, and uni-directional communications. The main idea of the proof is to show that this allocation strategy leads to a DoF constraint that constitutes no cycles. Details are provided in Appendix \ref{App:OptProof}. With this, the proof of achievability of Theorem~\ref{Thm:DoFKUsers} is complete.

\section{MIMO Y-channel with $N>M$}
\label{Sec:NgM}
The MIMO Y-channel has different DoF based on the relation between $M$ and $N$. The relation between the two can classified qualitatively into three regimes. One regime corresponds to the case where $N\geq KM$. The other corresponds to the case where $N\leq M$. In the intermediate regime, the problem becomes more challenging. From this point of view, it is important to study the former two regimes and explore their properties in order to come one step closer to a general solution.

In this paper, we have considered the regime where $M\geq N$. In this case, the columns of the uplink channel matrix $\H_i$ span the whole receive signal-space at the relay. Therefore, the spaces spanned by the columns of $\H_i$ and $\H_j$, $j\neq i$, completely overlap. Consequently, the users have to share this signal-space in an optimal manner in order to achieve the DoF of the channel. The optimal scheme has been developed in this paper, where the importance of cyclic communication has been demonstrated.

In the other regime corresponding to $N\geq KM$, the columns of the uplink matrix $\H_i$ span a sub-space of the receive signal-space at the relay, and the sub-spaces spanned by $\H_i$ and $\H_j$ do not intersect. Here the users do not have to share dimensions of the signal-space, as these dimensions are sufficiently many. Decode-forward becomes optimal in this case. In particular, the uplink is treated as a multiple access channel, and the downlink as a broadcast channel. The achievable DoF region is described by
\begin{align}
\sum_{j\in\mathcal{K}\setminus\{i\}} d_{ij}&\leq M,\\
\sum_{j\in\mathcal{K}\setminus\{i\}} d_{ji}&\leq M,
\end{align}
for all $i\in\mathcal{K}$, and it coincides with the cut-set bounds.

In the intermediate case where $M<N<KM$, the columns spanned by the channel matrices $\H_i$, $i=1,\cdots,K$, intersect at the relay. For instance, if $2M>N$, then each two users share $2M-N$ dimensions at the relay. If $2M\leq N$ and $3M>N$, then each three users share $3M-N$ dimensions at the relay, and so on. The main difference in this case is that pre-coding using the pseudo-inverse is not permissible. However, a similar scheme can be applied after designing appropriate pre- and post-coding matrices.

In a $K$-user Y-channel where the relay has $N>M$ antennas, the relay can use post-coding to recombine the received signals over each antenna to form $N$ observations, each with $KM-(N-1)$ variables. The result after this procedure is that the channel can be represented as multiple sub-channels, each shared by a subset of the $K$ users. Note that the relay has the freedom in choosing these observations judiciously. Consider the following example. Let the transmit signal of user $i$ be $\x_i=\V_i\u_i$ where $\u_i\in\mathbb{C}^{M}$ and $\V_i\in\mathbb{C}^{N\times M}$ is a beamforming matrix. The relay receives
\begin{align}
\y_r=\underbrace{[\H_1\V_1,\ \H_2\V_2,\ \cdots,\ \H_K\V_K]}_{\H}\u+\z_r,
\end{align}
where $\u=[\u_1^T,\ \u_2^T,\ \cdots,\ \u_K^T]^T$. Suppose that the relay wants to construct an observation involving the first $KM-(N-1)$ components of $\u$. The relay constructs $\hat{\y}_r$ as $\hat{\y}_r=\widehat{\H}^{-1}\y_r$ where $\widehat{\H}\in\mathbb{C}^{N\times N}$ is the matrix consisting of the last $N$ columns of $\H$. The first component of $\hat{\y}_r$ is a combination of the first $KM-(N-1)$ components of $\u$.

Note that in addition to this elimination of variables, some additional variables can be eliminated by the relay if they are aligned by the transmitters. In other words, if user $i$ sends the signal $\x_i=\sum_{k=1}^M \v_{ik}u_{ik}$ where $\v_{ik}\in\mathbb{C}^M$ is the $k$-th row of $\V_i$, so that $\H_i\v_{ik}=\H_j\v_{jk'}$ for some $i,j\in\mathcal{K}$ and some $k,k'\in\{1,\cdots,M\}$, then the signals $u_{ik}$ and $u_{jk'}$ align at the relay. In this case, eliminating $u_{ik}$ also eliminated $u_{jk'}$.

According to this discussion, the design of the optimal scheme is not an straightforward extension of the case considered in this paper. The main additional ingredient is the {\it design of the optimal pre-coders and post-coders for a given DoF tuple} so that the desired observations are obtained at the relay. We did not have to go through this step in this paper since for $N\leq M$, the same pre-coding allows achieving all DoF tuples. Given the pre- and post-coders, the coding schemes discussed in this paper (uni-direction, bi-directional, and cyclic) can be used over the resulting sub-channels. It is not clear whether such a combination would be optimal in general. The problem of designing the optimal scheme for $N\geq M$ thus remains an open problem. The sum DoF of the 4-user case has been characterized in \cite{ChenweiWang}.

It is worth to mention that the outer bound derived in this paper also applies for $N>M$. In general, the outer bound can be stated as
\begin{align}
\sum_{i=1}^{K-1}\sum_{j=i+1}^{K}d_{p_ip_j}\leq \min\{N,(K-1)M\}, \quad \forall \mathbf{p}
\end{align}
where $\mathbf{p}$ is a permutation of $(1,\cdots,K)$ and $p_i$ is its $i$-th component (see Appendix \ref{App:OptProof}). Combined with the the cut-set bounds
\begin{align}
\sum_{j\in\mathcal{K}\setminus\{i\}} d_{ij}&\leq\min\{M,N\},\\
\sum_{j\in\mathcal{K}\setminus\{i\}}^K d_{ji}&\leq\min\{M,N\},
\end{align}
for all $i\in\mathcal{K}$, we get a general outer bound. As discussed above, this outer bound is tight for $N\leq M$ and for $N\geq KM$. However, we expect that it is not tight in the intermediate regime. Similarly, the inner bound developed in this paper holds in for a general MIMO Y-channel as
\begin{align}
\sum_{i=1}^{K-1}\sum_{j=i+1}^{K}d_{p_ip_j}\leq \min\{M,N\}, \quad \forall \mathbf{p},
\end{align}
since if $N>M$, we can deactivate $N-M$ antennas at the relay and still apply our scheme. This inner bound is also not tight in general. In conclusion, the DoF region of the general MIMO Y-channel remains an open problem, and requires further investigation.

\section{Remarks on Channel Separability}
\label{Sec:SubChannelSep}
An interesting aspect of MIMO systems is their channel separability/inseparability. Separability of a MIMO channel means that independent coding on each sub-channel suffices to achieve the DoF of the channel. A MIMO point-to-point channel is an example of a separable MIMO channel. The main consequence of this separability is that the transmission can be optimized (in terms of achievable rate) using water-filling. The optimal scheme in this case consists of channel diagonalization, separate coding, plus power allocation. Inseparability on the other hand means that joint encoding over multiple sub-channels is necessary to achieve the DoF of the channel. In particular, in an inseparable channel, signals sent over different sub-channels are not always independent, and decoding is performed by considering multiple sub-channels jointly at the receiver. A MIMO interference channel is an example of an inseparable MIMO channel \cite{CadambeJafar_Inseperability}. The optimal scheme in such cases becomes more sophisticated. In general, the processing at the transmitters and the receivers of a separable channel is simpler compared to that of an inseparable channel. In this section, we make some remarks on channel separation of the Y-channel. 

\subsection{Inseparability in terms of DoF region}
We have seen that the optimal strategy that achieves the DoF region of our setup is a combination of bi-directional, cyclic, and uni-directional strategies. The resulting combination leads to coding over several sub-channels of the MIMO system. More precisely, the cyclic strategy with cycle length $\ell>2$ requires coding over $\ell-1$ sub-channels. Let us examine what would happen if one were to use a channel separation approach instead.

In the channel separation approach, there is no interaction between different sub-channels, and the signals transmitted over a sub-channels can be decoded by only observing this particular sub-channel. While this is not possible for cyclic strategies with cycle length $\ell>2$, this is possible for the bi-directional and the uni-directional strategies. So what would happen if we would rely only on those two strategies?

We have seen in Section \ref{Sec:BiDirStra3User} that using these two schemes only over a 3-user MIMO Y-channel with $N=M=3$ is not sufficient. Namely, the DoF tuple $\hat{\d}=(2,0,1,1,1,0)$ can not be achieved by this combination as shown in the example in Section \ref{Sec:BiDirStra3User}. Consequently, a channel separation approach is not optimal in the given scenario. Rather than channel separation, one has to code over several sub-channels by using the $3$-cyclic strategy to achieve the given DoF tuple. The same behaviour can be shown for a general $K$-user MIMO Y-channel with $N\leq M$. In conclusion, the MIMO Y-channel is in general not separable.

\subsection{Separability in terms of sum-DoF}
However, a channel separation approach is optimal in terms of sum-DoF. If we are not interested in the DoF trade-off between different DoF component (a trade-off which is reflected by the DoF region) but we are rather interested in the sum-DoF, then the bi-directional strategy (which can be applied in a channel separation approach) suffices. To show this, note that the DoF region $\mathcal{D}_K$ implies that the sum-DoF is given by
\begin{align}
d_\Sigma=2N.
\end{align}
This can be shown by summing up the upper bound corresponding to $\p=(1,2,\cdots,K)$ and the one corresponding to $\p=(K,K-1,\cdots,1)$ in Theorem \ref{Thm:DoFKUsers}. To achieve $2N$ DoF in total, the resources ($N$ sub-channels) can be distributed among the $|\mathcal{S}_2|$ $2$-cycles of the Y-channel in any desired manner. Then, each pair of users in a $2$-cycle exchange two signals (one signal in each direction) over each sub-channel assigned to this $2$-cycle. We have $N$ sub-channels in total, and thus, this strategy achieves $2N$ DoF. A simple allocation would be to serve one pair of users at a time, and to change the served pair of users in a round-robin fashion. Since we have $|\mathcal{S}_2|=\frac{K!}{2(K-2)!}=K(K-1)/2$ $2$-cycles in the $K$-user Y-channel, this round-robin technique would achieve $\frac{2N}{K(K-1)}$ DoF per message. Consequently, $d_{ij}=\frac{2N}{K(K-1)}$ for all $i\neq j$, for a sum-DoF of $2N$ which is the optimal sum-DoF. Note that this scheme is fair; it achieves a symmetric DoF tuple where all users  get the same DoF. In conclusion, the MIMO Y-channel is separable from sum-DoF point of view.

Note that throughout this work, the uplink and downlink of the Y-channel were considered separately. No adaptive coding has been used at the source nodes. In other words, the signals sent by the users in the uplink are independent of what they received in the downlink. This separation turns out to be optimal for our problem. This kind of separability first appeared in the context of the Gaussian two-way channel \cite{Han} where adaptive coding is not necessary, and separation is optimal from capacity point of view.

\section{Conclusion}
\label{Sec:Conclusion}
We have characterized the DoF region of the MIMO Y-channel with $K$ users, $N$ antennas at the relay, and $M\geq N$ antennas at the users. The DoF region is proved to be achievable by using channel diagonalization in addition to a combination of bi-directional, cyclic, and uni-directional communication strategies. Channel diagonalization decomposes the MIMO channel into $N$ parallel SISO sub-channels over which the cyclic and uni-directional strategies are performed. The bi-directional and cyclic strategies use compute-forward at the relay to deliver several linear combinations of different signals to the users, such that each user is able to extract his desired signals. In other words, the main ingredient of these strategies is physical-layer network-coding. The uni-directional strategy is based on decode-forward. This combination of strategies is optimized by using a simple resource allocation approach. Namely, we allocate resources (sub-channels) to different strategies based on their efficiency, starting with the most efficient and ending with the least efficient one. 

Although this optimal resource allocation solution is intuitive, it has an interesting property. In order to design an optimal scheme, we have to combine strategies with different efficiencies. In other words, relying on the strategy with highest efficiency (bi-directional strategy) is not enough. 

As a by-product, we conclude that the MIMO Y-channel can not be separated into disjoint parallel sub-channels without degrading its performance. In general, one has to code over multiple sub-channels to achieve the whole DoF region of the channel.

The approach used in this paper can be applied to derive the capacity region of $K$-user SISO Y-channels within a constant gap. To do this, the cyclic communication strategies should be applied to derive the capacity region of the linear deterministic Y-channel. Then, the results can be extended to the Gaussian case as in \cite{ChaabanSezgin_ISIT12_Y}. This is left for future work.

\section*{Acknowledgments}
The authors would like to express their gratefulness to Dr. Karlheinz Ochs (RUB, Germany) for the fruitful discussions.

\appendices

\section{Converse of Theorem \ref{Thm:DoFKUsers}}
\label{Sec:Converse}
In this section, we prove the converse of Theorem \ref{Thm:DoFKUsers}. We need to show that the DoF region of the MIMO Y-channel with $N\leq M$ is outer bounded by
\begin{align}
\label{DoFRegUpperBound}
\sum_{i=1}^{K-1}\sum_{j=i+1}^{K}d_{p_ip_j}\leq N, \quad \forall \mathbf{p}
\end{align}
where $\mathbf{p}$ is a permutation of $(1,\cdots,K)$ and $p_i$ is its $i$-th component. Let us consider the permutation $\p=(K,K-1,\cdots,1)$ and prove the upper bound \eqref{DoFRegUpperBound} holds for this particular permutation. We need to show that any achievable DoF tuple must satisfy
\begin{align}
\label{DoFRegUpperBoundp}
\sum_{i=2}^{K}\sum_{j=1}^{i-1}d_{ij}\leq N.
\end{align}

This bound is shown by using the genie-aided upper bound in \cite{ChaabanSezginAvestimehr_YC_SC}. Let us consider $\tau$ uses of the channel, and let us give $w_{ij}$, for all $j>i$ and $i>1$ to user 1 as side information. Let us also give $\y_i^{\tau}$, $i=2,\cdots,K-1$ to user 1 as side information.

Now, consider any achievable rate for the channel, for which every node can obtain its messages with an arbitrarily small probability of error. This means that, after $\tau$ channel uses, user 1 can decode $(w_{21},\cdots,w_{K1})$ from $\y_1^{\tau}$, and $(w_{12}, \cdots, w_{1K})$. After decoding its desired messages, user 1 combines its side information with the decoded messages to obtain $(\y_2^{n}, w_{21}, w_{23}, \cdots, w_{2K})$, which is the same observation as that of user 2. This makes user 1 able to decode $(w_{32},\cdots,w_{K2})$ since user 2 can decode them from the same observation. Similarly, after this step, user 1 has knowledge of the observation of user 3 and can use it to decoded $(w_{43},\cdots,w_{K3})$, and so on, until user 1 knows all messages in the network through side information and through decoding. 

Thus, user 1 knowing his own messages ($K-1$ messages) and the messages in the side information ($(K-2)(K-1)/2$ messages), and knowing his received signals $\y_1^\tau$, and the received signals of user 2 to $K-1$, can decode his desired messages ($K-1$ messages) and all remaining $(K-2)(K-1)/2$ messages. Using Fano's inequality \cite{CoverThomas}, and defining $\widehat{\W}_i=(W_{i+1,i},\cdots,W_{Ki})$ and $\W_i=(W_{i,i+1},\cdots,W_{iK})$ for $i=1,\cdots,K-1$, we can write\footnote{We drop the dependence of $R_{ij}$ on $\rho$ for clarity.}
\begin{align*}
\tau\left(\sum_{j=1}^{K-1}\sum_{i=j+1}^K R_{ij}-\varepsilon_\tau\right)&\leq I(\widehat{\W}_{1},\widehat{\W}_{2},\cdots,\widehat{\W}_{K-1};\vec{y}_1^\tau,\vec{y}_2^\tau,\cdots,\vec{y}_{K-1}^\tau,\W_1,\W_{2},\cdots,\W_{K-1})\nonumber\\
&\leq h(\vec{y}_1^\tau,\vec{y}_2^\tau,\cdots,\vec{y}_{K-1}^\tau)-h(\vec{y}_1^\tau,\vec{y}_2^\tau,\cdots,\vec{y}_{K-1}^\tau|\x_r^{\tau})\\
&= I(\x_r^{\tau};\vec{y}_1^\tau,\vec{y}_2^\tau,\cdots,\vec{y}_{K-1}^\tau)
\end{align*}
where $\varepsilon_\tau\to0$ as $\tau\to\infty$, and where the second step follows by using the definition of mutual information, the fact that conditioning does not increase entropy, and the Markov chain 
\begin{align*}
(\widehat{\W}_{1},\widehat{\W}_{2},\cdots,\widehat{\W}_{K-1},\W_1,\W_{2},\cdots,\W_{K-1})\to \x_r^\tau\to (\vec{y}_1^\tau,\vec{y}_2^\tau,\cdots,\vec{y}_{K-1}^\tau).
\end{align*}
We can write this bound as
\begin{align}
\tau\left(\sum_{j=1}^{K-1}\sum_{i=j+1}^K R_{ij}-\varepsilon_\tau\right)&\leq I(\x_r^{\tau};\D \x_r^{\tau}+\z^{\tau})
\end{align}
where
\begin{align}
\D=\begin{bmatrix}\D_1\\\D_2\\\vdots\\\D_{K-1}\end{bmatrix}\quad \text{and}\quad \z=\begin{bmatrix}\z_1\\\z_2\\\vdots\\\z_{K-1}\end{bmatrix}.
\end{align}
But this is the mutual information between the input $\x_r$ and the output $\D\x_r+\z$ of a MIMO $N\times (K-1)M$ point-to-point channel. This channel has $\min\{N,(K-1)M\}=N$ DoF \cite{Telatar}. Therefore, by dividing by $\tau$ and then letting $\tau\to\infty$ we get
\begin{align*}
\sum_{j=1}^{K-1}\sum_{i=j+1}^K R_{ij}\leq  N\log(\rho) +\mathcal{O}(1),
\end{align*}
which proves that 
\begin{align}
\sum_{j=1}^{K-1}\sum_{i=j+1}^K d_{ij}\leq  N,
\end{align}
which is equivalent to \eqref{DoFRegUpperBoundp}. This proves \eqref{DoFRegUpperBound} for the permutation $\p=(K,K-1,\cdots,1)$.  The upper bounds for all other permutations can be proved similarly. This concludes the proof of the converse of Theorem \ref{Thm:DoFKUsers} and shows the optimality of the diagonalization strategy, transmission strategies, and resource allocation strategy.

\section{DoF of Compute-forward}
\label{App:Rate2DoF}
\subsection{Uplink}
Consider two users 1 and 2 sending codewords $u_{1}^\tau$ and $u_{2}^\tau$, respectively, to a relay node. The codewords are constructed by using a nested-lattice code \cite{NazerGastpar} with power $P$ and rate $R$. In particular, both users uses a nested-lattice code with a shaping lattice $\Lambda$. User $i\in\{1,2\}$ constructs $u_i^\tau=\zeta_i[(t_i^\tau+d_i^\tau)\bmod\Lambda]$ and sends it, where $t_i^\tau$ is a codeword from the nested-lattice codebook, $d_i^\tau$ is a random dither (see \cite{NazerGastpar} for details), and $\zeta_i$ is a scaling parameter. The relay wants to decode a linear combination of $u_{1}^\tau$ and $u_{2}^\tau$. It receives
\begin{align}
y_r^\tau=h_1u_{1}^\tau+h_2u_{2}^\tau+z_r^\tau,
\end{align}
where $z_r^\tau$ is an i.i.d. $\mathcal{CN}(0,1)$. By choosing $\zeta_1$ and $\zeta_2$ so that $h_1\zeta_1=h_2\zeta_2=\min\{|h_1|,|h_2|\}$, the received codewords from users 1 and 2 align at the relay, and the relay can decode $(t_1^\tau+t_2^\tau)\bmod\Lambda$ as long as the rate of the code is bounded by \cite{NazerGastpar,ChaabanSezginAvestimehr_YC_SC}
\begin{align*}
R&\leq R^{ul}=\min\left\{
\log\left(\frac{1}{2}+|h_1|^2P\right),
\log\left(\frac{1}{2}+|h_2|^2P\right)
\right\}.
\end{align*}
The relay then is able to recover $h_1u_{1}^\tau+h_2u_{2}^\tau$ from $(t_1^\tau+t_2^\tau)\bmod\Lambda$ as shown in \cite{Nazer_IZS2012}. From DoF point of view, this process bounds the DoF of the signals sent by users 1 and 2 by one since $\lim_{P\to\infty}\frac{R^{ul}}{\log(P)}=1$. Thus, computation by the relay in the uplink leads to a DoF constraint of 1 DoF per stream.

\subsection{Downlink}
Now assume that the relay wants to send the sum $h_1u_{1}^\tau+h_2u_{2}^\tau$ to user 3 which also knows $u_2^\tau$ but wants to decode $u_1^\tau$. The relay sends $x_r^\tau=f(h_1u_{1}^\tau+h_2u_{2}^\tau)$ over the channel, where $f(\cdot)$ is an encoding function, and $x_r^\tau$ has power $P$. User 3 receives
\begin{align}
y_3^\tau=d_3x^\tau_r+z_3^\tau
\end{align}
where $z_r^\tau$ is an i.i.d. $\mathcal{CN}(0,1)$. Then, user 3 decodes $x_r^\tau$ from $y_3^\tau$ and uses its knowledge of $u_2^\tau$ to extract $u_1^\tau$ (broad-cast with side-information \cite{Tuncel,OechteringBjelakovicSchnurrBoche}. This decoding is possible if
\begin{align}
R\leq\log(1+|d_3|^2P)
\end{align}
From DoF point of view, this bounds the DoF of the first user's signal by one. Thus, decoding a compute-forward signal in the downlink leads to a DoF constraint of 1.

\section{Proof of Lemma \ref{Lem:Opt}}
\label{App:OptProof}
To prove the optimality of the proposed resource allocation, let us start by writing the number of sub-channels required to achieve a DoF tuple $\d\in\mathcal{D}_K$ by using the combination of bi-directional, $\ell$-cyclic, and uni-directional strategies with the resource allocation explained in Section \ref{Sec:ResAlloc}. The number of required sub-channels is given by
\begin{align}
\label{eq:Ns}
N_s&=\sum_{\ell=2}^K \sum_{m=1}^{|\mathcal{S}_\ell|} (\ell -1)d_{\c_{\ell [m]}} +\sum_{\e\in\mathcal{E}} d_{\e}^u.
\end{align}
The first summation in \eqref{eq:Ns} represents the number of sub-channels required by the bi-directional strategy and all the $\ell$-cycle strategies (an $\ell$-cycle strategy requires $(\ell-1)$ sub-channels as shown in Table \ref{Tab:Schemes}). The second sum represents the number of sub-channels required by the uni-directional strategy. Next, we substitute \eqref{eq:du} in \eqref{eq:Ns} to get

\begin{align}
\label{eq:Ns2}
N_s&=\sum_{\ell=2}^K \sum_{m=1}^{|\mathcal{S}_\ell|} (\ell-1)d_{\c_{\ell [m]}} +\sum_{\e\in\mathcal{E}} d_{\e}-\sum_{\e\in\mathcal{E}}\sum_{\ell=2}^{K}
\sum_{m=1}^{|\mathcal{S}_\ell|}\ind_{\mathcal{E}_{\c_{\ell [m]}}}(\e)d_{\c_{\ell [m]}}\\
&=\sum_{\ell=2}^K \sum_{m=1}^{|\mathcal{S}_\ell|} (\ell-1)d_{\c_{\ell [m]}} +\sum_{\e\in\mathcal{E}} d_{\e}-\sum_{\ell=2}^{K}
\sum_{m=1}^{|\mathcal{S}_\ell|}\sum_{\e\in\mathcal{E}}\ind_{\mathcal{E}_{\c_{\ell [m]}}}(\e)d_{\c_{\ell [m]}}\\
\label{eq:elledges}
&=\sum_{\ell=2}^K \sum_{m=1}^{|\mathcal{S}_\ell|} (\ell-1)d_{\c_{\ell [m]}} +\sum_{\e\in\mathcal{E}} d_{\e} -\sum_{\ell=2}^{K}
\sum_{m=1}^{|\mathcal{S}_\ell|}\ell d_{\c_{\ell [m]}}\\
\label{eq:Ns3}
&=\sum_{\e\in\mathcal{E}} d_{\e} -\sum_{\ell=2}^{K}
\sum_{m=1}^{|\mathcal{S}_\ell|} d_{\c_{\ell [m]}},
\end{align}
where \eqref{eq:elledges} follows since $\ell$ edges in $\mathcal{E}$ are edges of the cycle $\c_{\ell [m]}$. This is the required number of sub-channels for achieving $\d$ by our strategy. Since we have $N$ sub-channels in our Y-channel, we need the condition 
\begin{align}
N_s\leq N,
\end{align}
to hold for any $\d\in\mathcal{D}_K$. To show that $N_s\leq N$, we need to show that the MFG defined by the DoF components in \eqref{eq:Ns3} satisfies the no-cycle property. We denote this MFG by $\mathcal{G}$. The subtraction of the DoF of all cycles $d_{\c_{\ell [m]}}$ from $\sum_{\e\in\mathcal{E}} d_{\e}$ in \eqref{eq:Ns3} guarantees that $\mathcal{G}$ satisfies the no-cycle property as we show next.

\subsubsection{$\mathcal{G}$ has no $2$-cycles} All $2$-cycles $(i,j)$ in \eqref{eq:Ns3} are resolved by $-d_{(i,j)}$. To show this, we write
\begin{align}
\label{eq:NsN3}
N_s&=N_3 -\sum_{\ell=3}^{K}
\sum_{m=1}^{|\mathcal{S}_\ell|} d_{\c_{\ell [m]}},
\end{align}
where
\begin{align}
\label{EqN3ForMFG}
N_3=\sum_{\e\in\mathcal{E}} d_{\e} -\sum_{m=1}^{|\mathcal{S}_2|} d_{\c_{2 [m]}}.
\end{align}
The MFG defined by \eqref{EqN3ForMFG} might contain cycles of length 3 or more, but contains no $2$-cycles. Namely, since for $\c_{2[n]}=(i,j)$, we have $d_{\c_{2[n]}}=d_{(i,j)}=\min\{d_{ij},d_{ji}\}$, we get $d_{ij}+d_{ji}-d_{\c_{2[n]}}=\max\{d_{ij},d_{ji}\}$ \eqref{eq:DoF2Cycle}. 
This resolves the $2$-cycle $(i,j)$. Let us define the set $\mathcal{E}_3$ as the set of edges that remain after removing the edges $\arg\min_{\e\in\mathcal{E}_{\c_{2[n]}}}\{d_{\e}\}$, $n=1,\cdots,|\mathcal{S}_2|$, from $\mathcal{E}$. Thus,
$$\mathcal{E}_3=\mathcal{E}\setminus\bigcup_{n=1}^{|\mathcal{S}_2|} \left\{\arg\min_{\e\in\mathcal{E}_{\c_{2[n]}}}\ d_{\e}\right\}.$$
Clearly the set $\mathcal{E}_3$ has no $2$-cycles. Using this definition, we can write $N_3=\sum_{\e\in\mathcal{E}_3} d_{\e}$, and thus, we can write $N_s$ as
\begin{align}
\label{Eq:NsNo2Cycles}
N_s&=\sum_{\e\in\mathcal{E}_3} d_{\e} -\sum_{\ell=3}^{K}
\sum_{m=1}^{|\mathcal{S}_\ell|} d_{\c_{\ell [m]}}.
\end{align}
Next, we show that the terms $-d_{\c_{\ell [m]}}$ in \eqref{Eq:NsNo2Cycles} guarantee that the MFG defined by \eqref{Eq:NsNo2Cycles} has no $3$-cycles.

\subsubsection{$\mathcal{G}$ has no $3$-cycles}
The first sum in \eqref{Eq:NsNo2Cycles} might constitute $3$-cycles. However, if we write $N_s$ in \eqref{Eq:NsNo2Cycles} as
\begin{align}
\label{Eq:NsN4}
N_s&=N_4-\sum_{\ell=4}^{K}
 \sum_{m=1}^{|\mathcal{S}_\ell|} d_{\c_{\ell [m]}},
\end{align}
where 
\begin{align}
\label{eq:DoFsum3}
N_4=\sum_{i=1}^K\sum_{j=i+1}^K\max\{d_{ij},d_{ji}\}- \sum_{m=1}^{|\mathcal{S}_3|} d_{\c_{3 [m]}},
\end{align} 
we can show that the MFG described by \eqref{eq:DoFsum3} has no $3$-cycles (but possible cycles of length 4 or more). To this end, suppose that the first sum in \eqref{eq:DoFsum3} has a $3$-cycle $\c_{3[n]}=(i_1,i_2,i_3)$, i.e., the maximization in the first sum yields $d_{i_1i_2}$, $d_{i_2,i_3}$, and $d_{i_3i_1}$, and hence, $d_{\c_{3[n]}}$ defined as (cf. \eqref{eq:dc3m})
\begin{align}
\label{dc3nOpt}
d_{\c_{3[n]}}&=\min_{\e\in\mathcal{E}_{\c_{3[n]}}}
\left\{
d_{\e}
-\sum_{m=1}^{|\mathcal{S}_2|}\ind_{\mathcal{E}_{\c_{2[m]}}}(\e)d_{\c_{2[m]}}-\sum_{m=1}^{n-1}\ind_{\mathcal{E}_{\c_{3[m]}}}(\e)d_{\c_{3[m]}}
\right\},
\end{align}
is strictly positive. Further, assume that the minimization in \eqref{dc3nOpt} is achieved by the edge $i_1i_2$,\footnote{without loss of generality since we can always re-index the cycle accordingly if the minimization is achieved by another edge} i.e.,
\begin{align}
\label{dc3ni1i2}
d_{\c_{3[n]}}=d_{i_1i_2}-d_{(i_1,i_2)}
-\sum_{m=1}^{n-1}\ind_{\mathcal{E}_{\c_{3[m]}}}(i_1i_2)d_{\c_{3[m]}}.
\end{align}
Then, we can write \eqref{eq:DoFsum3} as
\begin{align}
\label{Eq:Subdc3ni1i21}
N_4&=\sum_{i=1}^K\sum_{j=i+1}^K\max\{d_{ij},d_{ji}\}-d_{\c_{3 [n]}}-\sum_{\substack{m=1\\m\neq n}}^{|\mathcal{S}_3|} d_{\c_{3 [m]}}\\
&=\sum_{i=1}^K\sum_{j=i+1}^K\max\{d_{ij},d_{ji}\}-\sum_{\e\in\mathcal{E}_{\c_{3[n]}}}d_{\e}
+\sum_{\e\in\mathcal{E}_{\c_{3[n]}}}d_{\e}
-d_{\c_{3 [n]}}
-\sum_{\substack{m=1\\m\neq n}}^{|\mathcal{S}_3|} d_{\c_{3 [m]}}\\
\label{Eq:Subdc3ni1i2}
&=\sum_{i=1}^K\sum_{j=i+1}^K\max\{d_{ij},d_{ji}\}-\sum_{\e\in\mathcal{E}_{\c_{3[n]}}}d_{\e}+d_{i_2i_3}+d_{i_3i_1}+d_{(i_1,i_2)}
+\sum_{m=1}^{n-1} \ind_{\mathcal{E}_{\c_{3[m]}}}(i_1i_2)d_{\c_{3[m]}}
-\sum_{\substack{m=1\\m\neq n}}^{|\mathcal{S}_3|} d_{\c_{3 [m]}}\\
\label{eq:N_s3CycleResolve}
&=\sum_{i=1}^K\sum_{j=i+1}^K\max\{d_{ij},d_{ji}\}
-\sum_{\e\in\mathcal{E}_{\c_{3[n]}}}d_{\e}
+d_{i_2i_3}+d_{i_3i_1}+d_{(i_1,i_2)}
-\sum_{m=1}^{n-1} \bar{\ind}_{\mathcal{E}_{\c_{3[m]}}}(i_1i_2)d_{\c_{3[m]}}-\sum_{m=n+1}^{|\mathcal{S}_3|} d_{\c_{3 [m]}},
\end{align}
where in \eqref{Eq:Subdc3ni1i2} we have substituted \eqref{dc3ni1i2}. Now, since $d_{c_{3[n]}}>0$, this implies that $d_{i_1i_2}\geq d_{i_2i_1}$ and hence $d_{(i_1,i_2)}=d_{i_2i_1}$. Substituting in \eqref{eq:N_s3CycleResolve}, we get
\begin{align}
 \label{eq:Ns4+cycles}
N_4&=\sum_{i=1}^K\sum_{j=i+1}^K\max\{d_{ij},d_{ji}\}
-\sum_{\e\in\mathcal{E}_{\c_{3[n]}}}d_{\e}+(d_{i_2i_1}+d_{i_2i_3}+d_{i_3i_1})
-\sum_{m=1}^{n-1} \bar{\ind}_{\mathcal{E}_{\c_{3[m]}}}(i_1i_2)d_{\c_{3[m]}}-\sum_{m=n}^{|\mathcal{S}_3|} d_{\c_{3 [m]}}.
\end{align}
As a result, the term $-d_{\c_{3[n]}}$ resolves the cycle $(i_1,i_2,i_3)$ by replacing $\sum_{\e\in\mathcal{E}_{\c_{3[n]}}}d_{\e}=d_{i_1i_2}+d_{i_2i_3}+d_{i_3i_1}$ with $d_{i_2i_1}+d_{i_2i_3}+d_{i_3i_1}$ which does not constitute a cycle. A similar procedure can be used to show that the term $-d_{\c_{3[n]}}$ resolves all $3$-cycles for $n=1,\cdots, |\mathcal{S}_3|$. For cycles $\c_{3[n]}$ which do not exist for the given $\d\in\mathcal{D}_K$, the corresponding $d_{\c_{3[n]}}$ is zero. As a result, the MFG defined by \eqref{eq:Ns4+cycles} contains neither $2$-cycles nor $3$-cycles. Thus, we can write $N_4$ as
\begin{align}
N_4=\sum_{\e\in\mathcal{E}_{4}}d_{\e},
\end{align}
where the set $\mathcal{E}_4$ is the set of edges that remain after removing the edges $\mathcal{F}_3(n)$, defined as
\begin{align*}
\mathcal{F}_3(n)=\arg\hspace{-.2cm}\min_{\e\in\mathcal{E}_{\c_{3[n]}}}
\hspace{-.1cm}\left\{
d_{\e}
-\sum_{m=1}^{|\mathcal{S}_2|}\ind_{\mathcal{E}_{\c_{2[m]}}}(\e)d_{\c_{2[m]}}
-\sum_{m=1}^{n-1}\ind_{\mathcal{E}_{\c_{3[m]}}}(\e)d_{\c_{3[m]}}
\hspace{-.1cm}\right\}
\end{align*}
(corresponding to \eqref{dc3nOpt}) for $n=1,\cdots,|\mathcal{S}_3|$, from $\mathcal{E}_3$. Thus,
\begin{align}
\mathcal{E}_4&=\mathcal{E}_3\setminus\bigcup_{n=1}^{|\mathcal{S}_3|} \mathcal{F}_3(n).
\end{align}
Clearly the set $\mathcal{E}_4$ has no $2$-cycles nor $3$-cycles, but possibly cycles of length 4 or more.\footnote{The set $\mathcal{E}_4$ is not fixed for all $\d\in\mathcal{D}_K$ since the remaining edges after resolving $2$-cycles and $3$-cycles depend on $\d$.} By substituting $N_4$ in $N_s$ \eqref{Eq:NsN4}, we can write
\begin{align}
\label{eq:Ns4Cycle}
N_s&=\sum_{\e\in\mathcal{E}_{4}}d_{\e}
-\sum_{\ell=4}^{K}
 \sum_{m=1}^{|\mathcal{S}_\ell|} d_{\c_{\ell [m]}}.
\end{align}
Now, it is obvious that $N_s$ has no $3$-cycles. Next, we show that it also has no $\ell$-cycles, $\ell=4,\cdots,K$.

\subsubsection{$\mathcal{G}$ has no cycles} We begin by writing $N_s$ in \eqref{eq:Ns4Cycle} as
\begin{align}
\label{eq:Ns4CycleN5}
N_s &=N_5-\sum_{\ell=5}^{K}
 \sum_{m=1}^{|\mathcal{S}_\ell|} d_{\c_{\ell[m]}},
\end{align}
where
\begin{align}
\label{eq:N5}
N_5&=\sum_{\e\in\mathcal{E}_{4}}d_{\e}
-\sum_{m=1}^{|\mathcal{S}_4|} d_{\c_{\ell [m]}}.
\end{align}
Again, we can show that the MFG defined by \eqref{eq:N5} does not contain $4$-cycles. In particular, suppose that the edges in $\mathcal{E}_4$ constitute the $4$-cycle $\c_{4[n]}=(i_1,i_2,i_3,i_4)$. Then, similar to above, assume that
\begin{align*}
d_{\c_{4[n]}}&=d_{i_1i_2}
-\sum_{\ell=2}^{3}
\sum_{m=1}^{|\mathcal{S}_\ell|}\ind_{\mathcal{E}_{\c_{\ell [m]}}}(i_1i_2)d_{\c_{\ell [m]}}-\sum_{m=1}^{n-1}\ind_{\mathcal{E}_{\c_{4[m]}}}(i_1i_2)d_{\c_{4[m]}}
\end{align*}
(cf. \eqref{eq:dcln}), and substitute in \eqref{eq:N5} to get
\begin{align*}
N_5&=\sum_{\e\in\mathcal{E}_{4}}d_{\e}
-\sum_{m=1}^{|\mathcal{S}_4|} d_{\c_{\ell [m]}}\\
&=\sum_{\e\in\mathcal{E}_{4}}d_{\e}
 -\hspace{-.1cm}\sum_{\e\in\mathcal{E}_{\c_{4[n]}}}d_{\e}
 +\hspace{-.1cm}\sum_{\e\in\mathcal{E}_{\c_{4[n]}}}d_{\e}
 -d_{\c_{4[n]}}-\sum_{\substack{m=1\\m\neq n}}^{|\mathcal{S}_4|} d_{\c_{4 [m]}}\\
&=\sum_{\e\in\mathcal{E}_{4}}d_{\e}
-\hspace{-.1cm}\sum_{\e\in\mathcal{E}_{\c_{4[n]}}}d_{\e}
 +(d_{i_2i_1}+d_{i_2i_3}+d_{i_3i_4}+d_{i_4i_1})-\sum_{m=1}^{n-1} \bar{\ind}_{\mathcal{E}_{\c_{4[m]}}}(i_1i_2)d_{\c_{4[m]}}-\sum_{m=n+1}^{|\mathcal{S}_4|} d_{\c_{4 [m]}},
\end{align*}
similar to \eqref{Eq:Subdc3ni1i21}-\eqref{eq:N_s3CycleResolve}, thus resolving this $4$-cycle by replacing $\sum_{\e\in\mathcal{E}_{\c_{4[n]}}}d_{\e}=d_{i_1i_2}+d_{i_2i_3}+d_{i_3i_4}+d_{i_4i_1}$ by $d_{i_2i_1}+d_{i_2i_3}+d_{i_3i_4}+d_{i_4i_1}$. Similarly, all $4$-cycles are resolved by the terms $-d_{\c_{4[m]}}$ leading to
\begin{align}
N_5=\sum_{\e\in\mathcal{E}_{5}}d_{\e},
\end{align}
where $\mathcal{E}_5$ is defined similar to $\mathcal{E}_4$, i.e.,
\begin{align*}
\mathcal{E}_5=\mathcal{E}_4\setminus\bigcup_{n=1}^{|\mathcal{S}_4|} \mathcal{F}_4(n),
\end{align*}
and
\begin{align*}
&\mathcal{F}_4(n)=\arg\min_{\e\in\mathcal{E}_{\c_{4 [n]}}}
\left\{
d_{\e}
-\sum_{i=2}^{3}
\sum_{m=1}^{|\mathcal{S}_i|}\ind_{\mathcal{E}_{\c_{i[m]}}}(\e)d_{\c_{i[m]}}-\sum_{m=1}^{n-1}\ind_{\mathcal{E}_{\c_{4 [m]}}}(\e)d_{\c_{4 [m]}}
\right\}.
\end{align*}

The edges of $\mathcal{E}_5$ do not constitute $2$-, $3$-, or $4$-cycles, but might constitute cycles of length 5 or more. By substituting $N_5$ in $N_s$ in \eqref{eq:Ns4CycleN5}, we get 
\begin{align}
\label{eq:Ns5Cycle}
N_s&=\sum_{\e\in\mathcal{E}_{5}}d_{\e}
-\sum_{\ell=5}^{K}
 \sum_{m=1}^{|\mathcal{S}_\ell|} d_{\c_{\ell [m]}}.
\end{align}
By proceeding similarly, we can show that all $\ell$-cycles in \eqref{eq:Ns5Cycle} are resolved, and that $N_s$ can be written as
\begin{align}
\label{eq:NsNoCycle}
N_s&=\sum_{\e\in\mathcal{E}_{K+1}}d_{\e}
\end{align}
where $\mathcal{E}_{K+1}$ is a set of edges that constitute no cycles of length $2,\cdots,K$. We conclude that $N_s$ is the sum of DoF components of $\d\in\mathcal{D}_K$ whose corresponding MFG satisfies the no-cycle property. Since \eqref{eq:DoFRegionKUser} implies that the sum of all permutations of $K(K-1)/2$ components of $\d$ constituting no cycles is less than $N$ for all DoF tuples $\d\in\mathcal{D}_K$, then $N_s\leq N$ by \eqref{eq:NsNoCycle}, which proves the achievability of $\mathcal{D}_K$.

\end{document}